\documentclass[%
reprint,
superscriptaddress,
notitlepage,
amsmath,amssymb,
aps,
onecolumn,
floatfix,
tightenlines,
longbibliography,
11pt,
]{revtex4-1}
\usepackage{psfrag,graphicx,epsfig,color}% Include figure files
\usepackage{dcolumn}% Align table columns on decimal point
\usepackage{bm}% bold math
\usepackage{natbib}
\usepackage{float}
\usepackage[usenames,dvipsnames,svgnames,table]{xcolor}
\usepackage{subfigure}

\def\re    {{R_\lambda}}
\def\uu {{\mathbf{u}}}
\def\ww {{\boldsymbol{\omega}}}

\def\ei {{\mathbf{e}_i}}

\begin{document}

\title{
%Vorticity amplification in high Reynolds number turbulence \\
%Vortex stretching and enstrophy production in high Reynolds number turbulence  
Generation of intense dissipation in high Reynolds number turbulence
}

%%% Authors %%%
\author{Dhawal Buaria }
\email[]{dhawal.buaria@nyu.edu}
%\thanks{}
\affiliation{Tandon School of Engineering, New York University, New York, NY 11201, USA}
\affiliation{Max Planck Institute for Dynamics and Self-Organization, 37077 G\"ottingen, Germany}
\author{Alain Pumir}
\affiliation{Laboratoire de Physique, ENS de Lyon, Universit\'e de Lyon 1 and CNRS, 69007 Lyon, France}
\affiliation{Max Planck Institute for Dynamics and Self-Organization, 37077 G\"ottingen, Germany}
\author{Eberhard Bodenschatz}
\affiliation{Max Planck Institute for Dynamics and Self-Organization, 37077 G\"ottingen, Germany}
\affiliation{Institute for Nonlinear Dynamics, University of G\"ottingen, 37077 G\"ottingen, Germany}

\date{\today}% It is always \today, today,

%\thispagestyle{empty}

\begin{abstract}

Intense fluctuations of energy dissipation rate
in turbulent flows
result from the self-amplification of strain rate
via a quadratic nonlinearity,
with contributions from vorticity (via the vortex
stretching mechanism) and pressure-Hessian --
which are analyzed here using
direct numerical simulations of isotropic
turbulence on
up to $12288^3$ grid points,
and Taylor-scale Reynolds numbers in the range $140-1300$.
We extract the statistics of various terms 
involved in the
amplification of strain and
condition them on the magnitude of strain.
We find that strain is self-amplified by
the quadratic nonlinearity,
and depleted via vortex stretching;
whereas pressure-Hessian acts to
redistribute strain fluctuations towards the mean-field
and hence depletes intense strain.
Analyzing the intense fluctuations of strain in
terms of its eigenvalues reveals that the net amplification
is solely produced by the third eigenvalue, resulting
in strong compressive action. In contrast, the self-amplification
acts to deplete the other two eigenvalues, whereas vortex
stretching acts to amplify them, with both effects canceling
each other almost perfectly.
The effect of the pressure-Hessian for each eigenvalue
is qualitatively similar to that of vortex stretching,
but significantly weaker in magnitude.
Our results conform with the familiar notion
that intense strain is organized in sheet-like structures,
which are in the vicinity of, but never overlap with
tube-like regions of intense vorticity due to
fundamental differences in their amplifying mechanisms.

\end{abstract}

\maketitle

%%%%%%%%%%%%%%%%%%%%%%%%%%%

\section{Introduction}

The dissipation rate of kinetic energy,
$\epsilon$, defined as:
\begin{align}
\epsilon = 2\nu S_{ij} S_{ij} \ , \ \ \text{where}  \ \ 
S_{ij} = \frac{1}{2} \left( \frac{\partial u_i}{\partial x_j} 
+ \frac{\partial u_j}{\partial x_i} \right) \ ,
\label{eq:diss}
\end{align}
plays an indispensable role in our understanding
of turbulent fluid flows.
Here, $\nu$ is the kinematic viscosity and
$S_{ij}$ is the strain rate tensor
(the symmetric part of the velocity gradient
tensor $\partial u_i/\partial x_j$).
The mean of dissipation rate quantifies
the net cascade of energy from large to small scales,
manifestly becoming 
independent of $\nu$, as $\nu\to0$
\cite{Frisch95,sreeni84,kaneda03}.
This property, also known as dissipative
anomaly, is the central tenet of nearly all turbulence theories
and models \cite{Frisch95}.
However, the fluctuations of dissipation rate
(and hence that of strain rate)
can be orders of magnitude larger than its mean \cite{MS91,BPBY2019},
a phenomena known as intermittency,
which renders any mean-field description of turbulence
inadequate \cite{Frisch95,Sreeni97}.
Understanding the formation of
such intense fluctuations
and characterizing their statistical properties
has long remained one of the outstanding challenges in
turbulence \cite{Frisch95, Tsi2009}.

Understanding the intense fluctuations of dissipation
is also directly important from a practical standpoint.
For instance, strong strain rates can greatly enhance
dispersion of particles and
influence mixing of scalars 
or can adversely affect flame propagation
in reacting flows \cite{BSY.2015, BCSY2021a, Pitsch2000, ham_pof11}. 
Intense strain also leads to generation
of intense vorticity, via the well-known vortex
stretching mechanism \cite{tl72}, which in turn 
influences clustering of inertial particles \cite{collins97}.
In fact, strain and vorticity are not independent
and their coupling implicitly encodes
all the multiscale interactions in the flow 
\cite{ham_pof08, BPB2020, BP2021}. 
While much attention has been recently given 
to understand this interaction in light of vorticity
amplification \cite{BBP2020, BPB2020, BP2021}
and energy cascade across scales \cite{carbone20, johnson21},
in the current work, we present a complementary
investigation focusing on amplification of strain
(and hence dissipation rate).

The key mechanisms controlling amplification of strain can be 
readily identified by writing its transport equation
(as derived from the incompressible Navier-Stokes equations):
\begin{align}
\frac{D S_{ij}}{D t} = 
-S_{ik} S_{kj} - 
\frac{1}{4} (\omega_i \omega_j - \omega_k \omega_k \delta_{ij}) 
- \Pi_{ij}
+ \nu \nabla^2 S_{ij}
\label{eq:dsdt}
\end{align}
where $\ww = \nabla \times \uu$ is the vorticity vector and 
$\Pi_{ij} = \frac{1}{\rho}\frac{\partial^2 P}{\partial x_i \partial x_j}$
is the pressure Hessian tensor.
The first term on the r.h.s. of Eq.~\eqref{eq:dsdt}
captures the self-amplification of strain,
which by itself
could lead to a finite time singularity.
The second term captures the influence of vorticity 
and essentially the feedback of vortex stretching 
on strain itself. The third term involving pressure-Hessian
represents the influence of non-local effects via the pressure
field, and hence couples the entire state of the flow.
This nonlocal dependence is readily seen by taking
the trace of Eq.~\eqref{eq:dsdt}, leading to the Poisson equation:
\begin{align}
\Pi_{ii} = \nabla^2 P/\rho = (\omega_i \omega_i - 2S_{ij} S_{ij})/2 \ .
\label{eq:poisson}
\end{align}
The final (linear) term in Eq.~\eqref{eq:dsdt} represents the viscous diffusion
of strain.

In this work, our main goal is to investigate various amplification
mechanisms leading to the formation intense strain 
and hence dissipation. To this end,
we analyze the statistics of the (inviscid) nonlinear
terms in Eq.~\eqref{eq:dsdt},
in particular by conditioning them on magnitude of strain.
One of the implicit goals is to also identify
and understand which (inviscid) mechanism(s) 
possibly contribute 
in preventing an unbounded growth of strain
\cite{BPB2020}.
We utilize data from high-resolution 
direct numerical simulations (DNS) 
of isotropic turbulence in periodic domains, 
which is the most efficient numerical tool to study the 
small-scale properties of turbulence.
Another important purpose
of the current study is also to understand the effect of increasing 
Reynolds number. With that in mind, we utilize a massive DNS database 
with Taylor-scale Reynolds number $\re$ ranging from 140
to 1300 on grid sizes going up to  $12288^3$,
with particular attention
on resolving the small-scales and hence 
the extreme fluctuations accurately \cite{BBP2020, BPB2020, BP2021}.

%In homogeneous isotropic turbulence (HIT), the contribution of strain 
%amplification from the strain quadratic nonlinearity
%is strongly reduced by the vorticity term, while the pressure term
%does not contribute~\cite{Tsi2009}. 
To get insight on the formation of intense strain 
we compute various statistics related
to strain amplification, conditioned on magnitude of strain. 
We find that the self-amplification solely drives the growth
of intense fluctuations, whereas vortex stretching and pressure
Hessian terms act to attenuate this growth.
By decomposing various contributions in the eigenframe of strain
tensor, we further show that this amplification
and attenuation predominantly occurs for the
most negative eigenvalue, signifying intense strain events
correspond to strong compressive motion.
In contrast, the other two eigenvalues 
are amplified by the vortex stretching mechanism and depleted
by the self-amplification term, with both these
mechanisms canceling each other almost perfectly.
The effect of the pressure Hessian, qualitatively similar
to that of vortex stretching, is to weakly
amplify these two eigenvalues.
The structure of the nonlinearities discussed 
here is consistent with the notion that regions of 
intense strain are organized in sheet-like 
structures \cite{moisy04, elsinga17},
which are unlikely to be colocated with regions
of intense vorticity organized in tube-like structures
\cite{Jimenez93,moffatt94} -- underscoring
the importance of non-local interactions between
strain and vorticity in amplifying gradients \cite{BPB2020,BP2021}.

The rest of the manuscript is organized as follows. 
In \S~\ref{sec:num}, we
briefly provide the details pertaining to DNS database utilized 
in this work.
The various nonlinearities controlling
the amplification of strain are investigated 
in \S~\ref{sec:stat_strain}, in particular
by analyzing their statistics on magnitude
of strain. 
In \S~\ref{sec:strain_decomp},
the various contributions are further analyzed in
the eigenbasis of strain tensor.
Finally, we summarize our results
in \S~\ref{sec:concl}.

%\cite{Ashurst87,Tsi2009,buxton10,Meneveau11,zhou16} --

\section{Numerical approach and database} 
\label{sec:num}

The data utilized here are the same
as in recent works 
\cite{BPBY2019, BS2020, BBP2020, BPB2020, BP2021}
and are generated using
direct numerical simulations (DNS) of incompressible
Navier-Stokes equations,
for the canonical setup 
of isotropic turbulence in a periodic
domain. The simulations are carried out
using highly accurate Fourier pseudo-spectral methods
with second-order Runge-Kutta integration in time,
and the large scales are forced numerically 
to achieve statistical stationarity. 
A key characteristic  of our data is that we have achieved
a wide range of Taylor-scale Reynolds number $\re$,
going from $140-1300$, while maintaining excellent small-scale
resolution on grid sizes of going up to $12288^3$.
The resolution is as high as 
$k_{\rm max} \eta \approx 6$, where
$k_{\rm max} = \sqrt{2}N/3$, is the maximum
resolved wavenumber on a $N^3$ grid, and $\eta$
is the Kolmogorov length scale.
Convergence with respect to resolution and statistical sampling has been
adequately established in previous works \cite{BBP2020, BPB2020}.
We summarize the DNS database and the simulation 
parameters in Table~\ref{tab:param}.

\begin{table}[h]
\centering
    \begin{tabular}{cccccc}
\hline
    $\re$   & $N^3$    & $k_{max}\eta$ & $T_E/\tau_K$ & $T_{sim}$ & $N_s$  \\
\hline
    140 & $1024^3$ & 5.82 & 16.0 & 6.5$T_E$ &  24 \\
    240 & $2048^3$ & 5.70 & 30.3 & 6.0$T_E$ &  24 \\
    390 & $4096^3$ & 5.81 & 48.4 & 2.8$T_E$ &  35 \\
    650 & $8192^3$ & 5.65 & 74.4 & 2.0$T_E$ &  40 \\
   1300 & $12288^3$ & 2.95 & 147.4 & 20$\tau_K$ &  18 \\
\hline
    \end{tabular}
\caption{Simulation parameters for the DNS runs
used in the current work: 
the Taylor-scale Reynolds number ($\re$),
the number of grid points ($N^3$),
spatial resolution ($k_{max}\eta$), 
ratio of large-eddy turnover time ($T_E$)
to Kolmogorov time scale ($\tau_K$),
length of simulation ($T_{sim}$) in statistically stationary state
and the number of instantaneous snapshots ($N_s$) 
used for each run to obtain the statistics.
}
\label{tab:param}
\end{table}

\section{Statistics conditioned on strain/dissipation}
\label{sec:stat_strain}

In order to quantify the intensity of strain,
we consider the quantity $\Sigma$ defined as 
\begin{align}
\Sigma = 2 S_{ij} S_{ij} 
\label{eq:def_Sigma}
\end{align}
which is simply the dissipation 
rate without the viscosity, i.e., $\Sigma=\epsilon/\nu$.
The benefit of directly using $\Sigma$ is that its mean
defines the Kolmogorov time scale $\tau_K$, i.e., 
$\langle \Sigma \rangle = 1/\tau_K^2$
and from homogeneity is also equal to the mean
of enstrophy, i.e., 
$\langle \Sigma \rangle = \langle \Omega \rangle$,
where $\Omega= \omega_i \omega_i$.
From Eq.~\eqref{eq:dsdt}, the following transport
equation for $\Sigma$ can be derived:
\begin{align}
\frac{1}{4} \frac{D \Sigma}{Dt}  = - S_{ij} S_{jk} S_{ki} 
- \frac{1}{4} \omega_i \omega_j S_{ij} - S_{ij} \Pi_{ij} + \nu S_{ij} \nabla^2 S_{ij} \ ,
\label{eq:ds2dt}
\end{align}
%{The first term on the r.h.s. of Eq.~\eqref{eq:ds2dt} expresses the strain
%self-amplification. The }
where the term $\omega_i \omega_j S_{ij}$,
which leads to production of enstrophy
when considering vorticity transport equation \cite{BBP2020},
clearly demonstrates the feedback of vortex
stretching on amplification of strain.
%Since $\omega_i \omega_j S_{ij}$ is known to be
%positive on average \cite{Betchov56}, it can be expected
%that regions of intense vorticity lead to depletion
%of strain.

In statistically stationary isotropic turbulence,
as considered in this work, the mean of the 
l.h.s. of Eq.~\eqref{eq:ds2dt} is zero.
For the terms on r.h.s., it is known that \cite{Tsi2009,Betchov56}:
\begin{align}
\langle S_{ij} \Pi_{ij} \rangle &= 0   \\ 
- \langle S_{ij} S_{jk} S_{kj} \rangle &= \frac{3}{4} \langle \omega_i \omega_j S_{ij} \rangle
\label{eq:Betchov}
\end{align}
Utilizing these relations, the averaging of Eq.~\eqref{eq:ds2dt} leads to:
\begin{align}
\langle S_{ij} S_{jk} S_{ki} \rangle = \frac{3}{2} \nu \langle S_{ij} \nabla^2 S_{ij} \rangle
\label{eq:Betchov_visc}
\end{align}
which gives a simple balance between inviscid production and 
viscous dissipation of strain.
Since $\langle \omega_i \omega_j S_{ij} \rangle$
is known to be positive on average~\cite{Betchov56},
%(Note, since $\langle\omega_i \omega_j S_{ij}\rangle>0$,
it follows that $- \langle S_{ij} S_{jk} S_{kj} \rangle >0$, 
therefore implying generation of strain via a self-amplification
mechanism.
While the results in Eqs.~\eqref{eq:Betchov}-\eqref{eq:Betchov_visc},
hold on average for the entire flow field, 
they do not imply any particular
relation when considering the same statistics 
conditioned on $\Sigma$
(which is required to isolate the extreme events
from the mean-field). 
In the following, 
we investigate
the role of various terms in Eq.~\eqref{eq:ds2dt} 
conditioned on $\Sigma$.

\subsection{Strain self-amplification and vortex stretching}
\label{subsec:strain_ampl}

\begin{figure}[h]
\begin{center}
\includegraphics[width=0.46\textwidth]{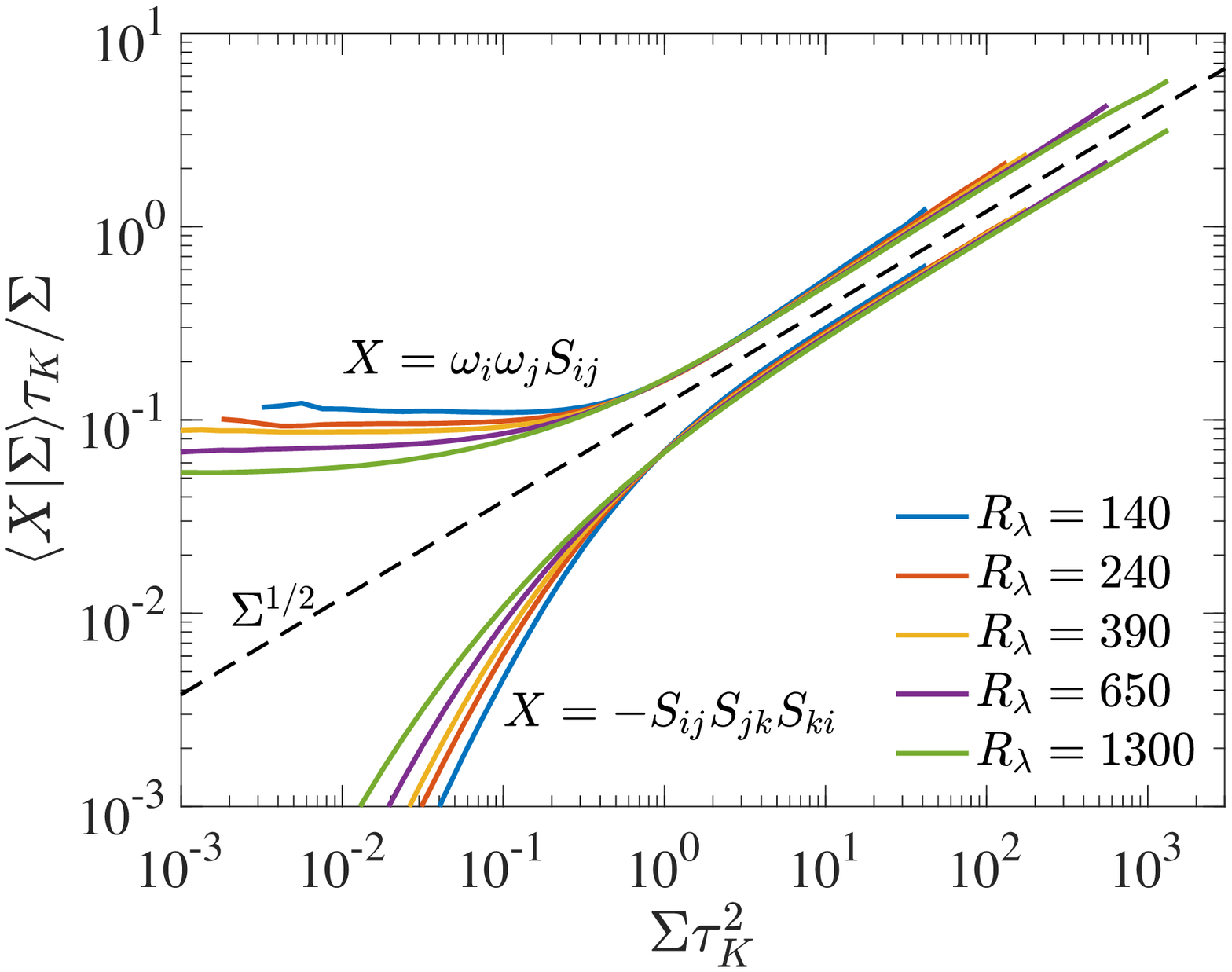} \ \ \ \ \ 
\includegraphics[width=0.46\textwidth]{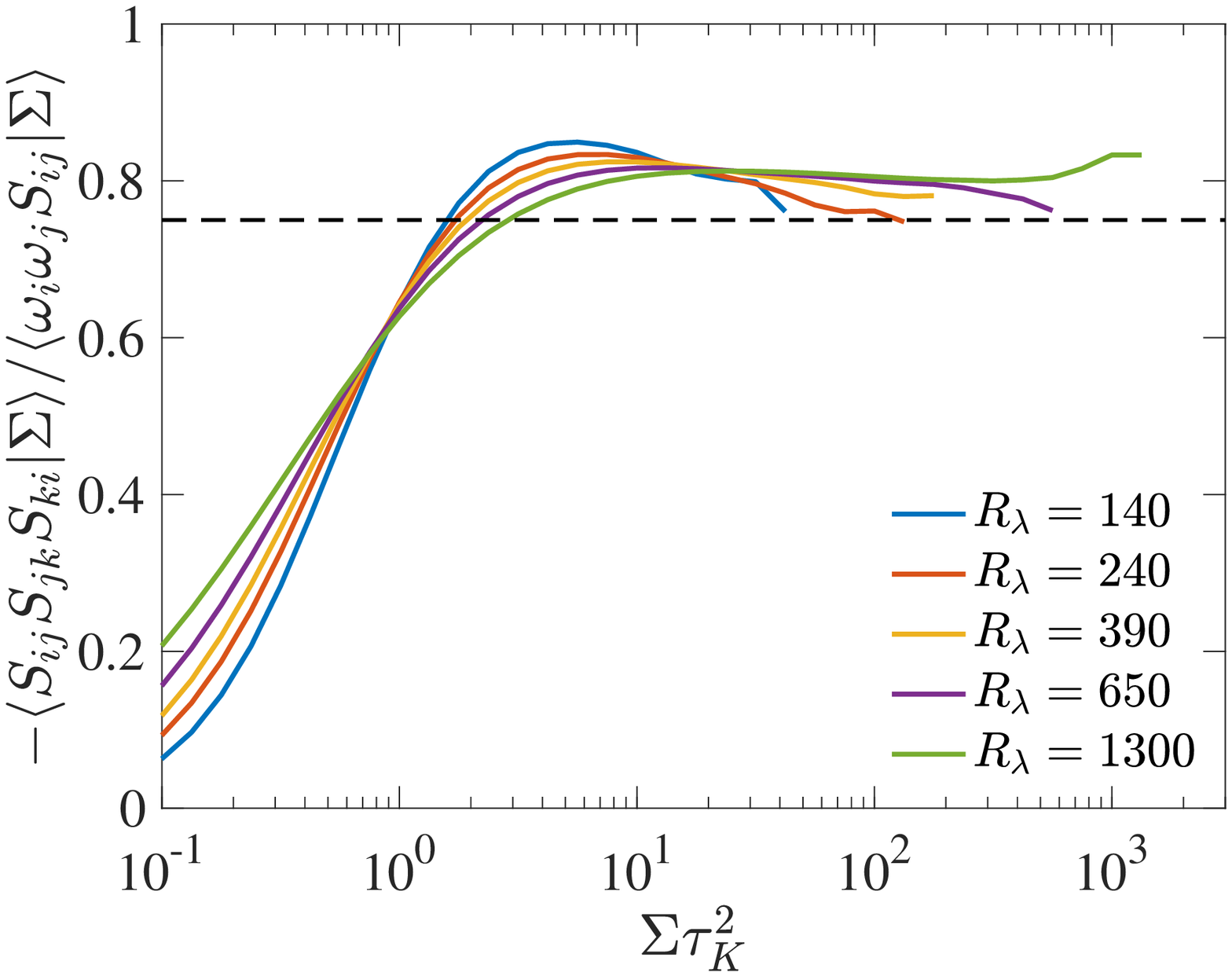} 
\caption{
(a) Conditional expectations (given $\Sigma$) 
of the strain self-amplification
and vortex stretching terms,
for various $\re$.
All quantities are normalized by Kolmogorov time scale 
$\tau_K = \langle \Sigma \rangle^{-1/2}$.
The black dashed line indicates the power-law $\Sigma^{1/2}$.
(b) The ratio of conditional strain self-amplification
and vortex stretching terms.  
The horizontal dashed line indicates the value $3/4$, 
corresponding to the ratio of unconditional averages
as given by Eq.~\eqref{eq:Betchov}.
}
\label{fig:trs3-str}
\end{center}
\end{figure}

In this subsection, we analyze the contributions from the 
self-amplification and vortex stretching terms
when conditioned on magnitude of strain, i.e.,
respectively $-\langle S_{ij}S_{jk} S_{ki} |\Sigma\rangle$
and $\langle \omega_i \omega_j S_{ij} |\Sigma \rangle$
Figure~\ref{fig:trs3-str}a
shows both
terms, divided by $\Sigma$ for convenience
and for various $\re$.
All quantities are appropriately non-dimensionalized
by $\tau_K$, which allows us to 
demarcate the strength of events with respect to the mean-field.
The main observation is that both the plotted quantities
in Fig.~\ref{fig:trs3-str}a scale as $\Sigma^{1/2}$
(marked by black dashed line) 
for events stronger than the mean ($\Sigma \tau_K^2 \gtrsim 1$), 
which implies the conditional expectations 
themselves scale as $\Sigma^{3/2}$ --
consistent with simple dimensional argument. 
Moreover, the dependence on $\re$ is very weak
(especially as $\re$ increases),
suggesting an asymptotic state has likely been reached.

It should be noted that the magnitude of the
vortex stretching term (which depletes strain) is larger,
but its net contribution is still lower than the self-amplification
term due to the factor of $1/4$ 
in Eq.~\eqref{eq:ds2dt}.
To further investigate their relative contributions,
Fig.~\ref{fig:trs3-str}b shows the ratio
of their conditional expectations for various $\re$. 
For extreme events of strain, 
we observe that the ratio 
seemingly asymptotes to 
a constant value of about $0.8$, 
which is different and slightly larger than  $3/4$,
the value  of their unconditional averages.
This implies that the 
%depletion
%{reduction of the strain self-amplification by vortex
%stretching is slightly weaker for $\Sigma \tau_K^2 > 1$ than the overall
%reduction, implied by Eqs.~\eqref{eq:ds2dt} and \eqref{eq:Betchov}.}
overall (negative) contribution of the 
vortex stretching term is 
is about $1/(4 \times 0.8) \approx 1/3.2$ times that of the 
strain self-amplification term, which
is slightly smaller than the factor $1/3$
valid for the overall field
(as seen from
Eqs.~\eqref{eq:ds2dt} and \eqref{eq:Betchov}).

\begin{figure}
\begin{center}
\includegraphics[width=0.46\textwidth]{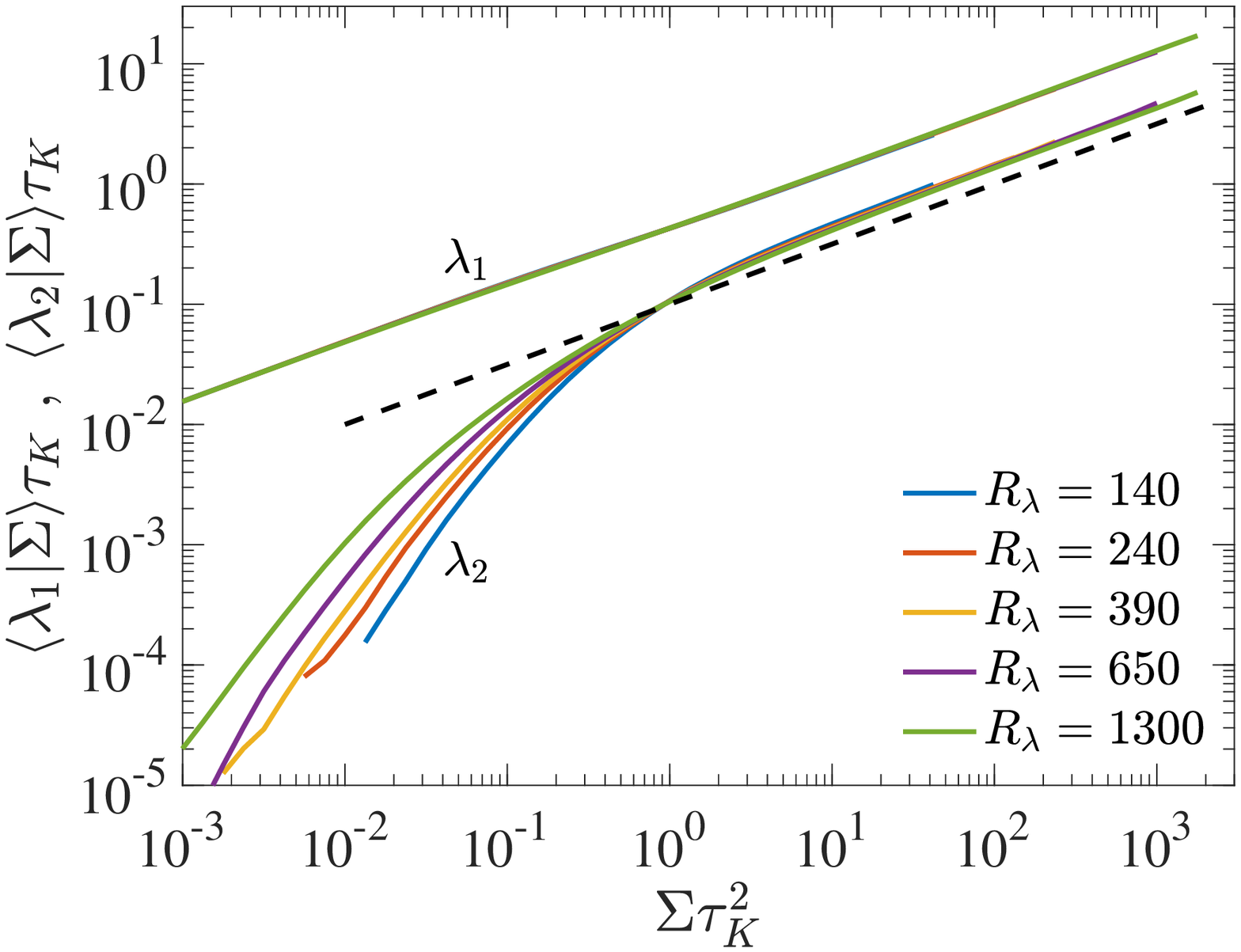} \ \ \ \ \ 
\includegraphics[width=0.46\textwidth]{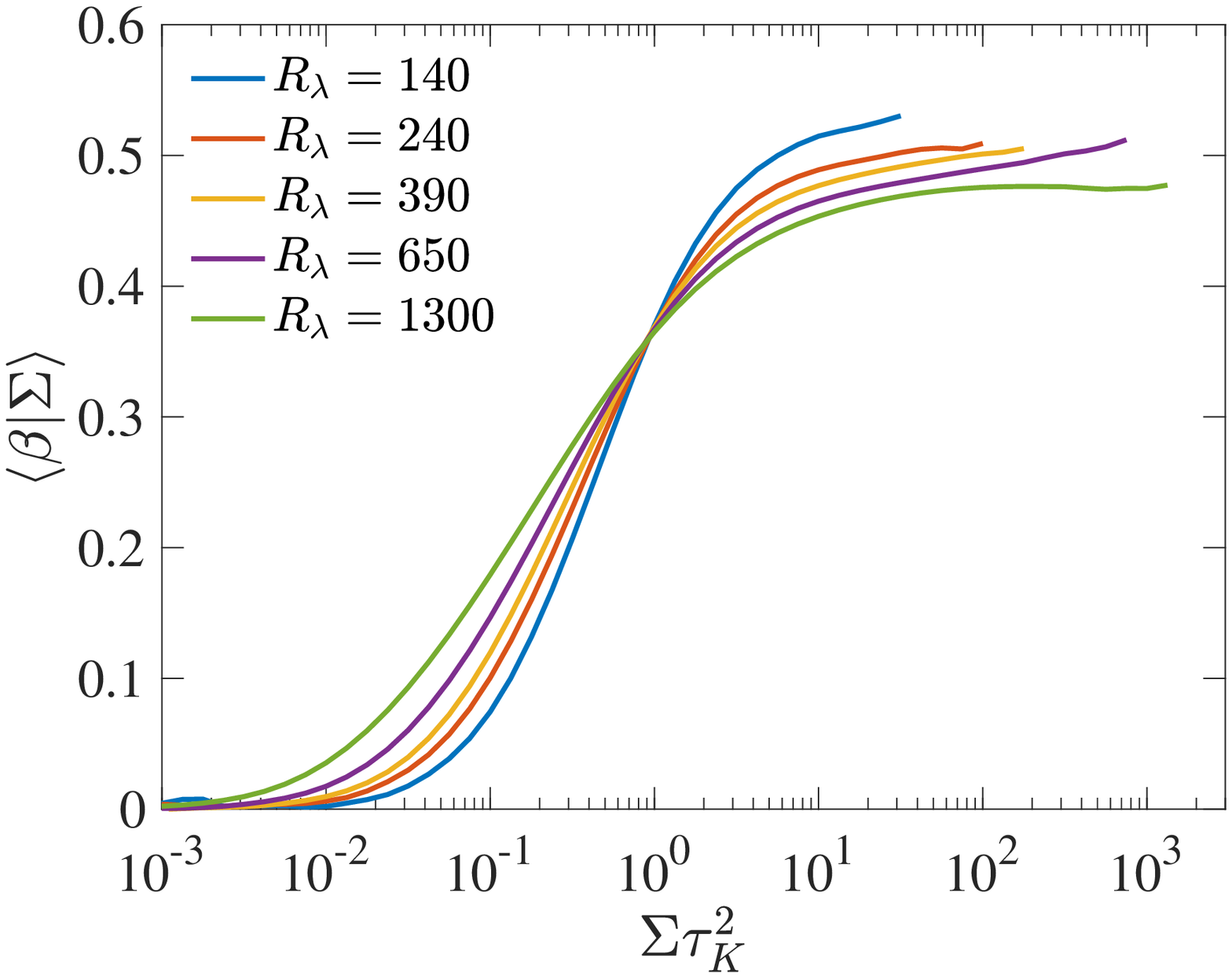}
\caption{
Conditional expectations (given $\Sigma$) of
(a) first two eigenvalues of
strain tensor, 
(b) the ratio $\beta$ as defined by Eq.~\eqref{eq:def_beta},
for various $\re$.
The dashed line in (a) indicates a power law of $\Sigma^{1/2}$.
}
\label{fig:lam12_diss}
\end{center}
\end{figure}

To better understand the results 
in Fig.~\ref{fig:trs3-str}, 
we analyze them further 
in the eigenframe 
of strain, as given by its
three eigenvalues $\lambda_i$ (for $i=1,2,3$, 
(such that $\lambda_1 \ge \lambda_2 \ge \lambda_3$)
and the corresponding eigenvectors $\mathbf{e}_i$.
Incompressibility imposes
$\lambda_1 + \lambda_2 + \lambda_3 = 0$,
which renders $\lambda_1$ to be always
positive (stretching) and $\lambda_3$ to be always negative
(compressive).
It is well known that the second (intermediate) eigenvalue
$\lambda_2$ is positive on average, leading
to net production of enstrophy \cite{Betchov56, Ashurst87, BBP2020}.
Using the eigenframe, we can readily show that
\begin{align}
\Sigma = 2 (\lambda_1^2 + \lambda_2^2 + \lambda_3^2) \ , \ \ \ \ \ 
S_{ij} S_{jk} S_{ki} = \lambda_1^3 + \lambda_2^3 + \lambda_3^3 
= 3 \lambda_1 \lambda_2 \lambda_3 \ . 
\end{align}

The power-law behavior of 
$ - \langle S_{ij} S_{jk} S_{ki} | \Sigma \rangle \sim \Sigma^{3/2}$
(for $\Sigma \tau_K^2 > 1$) suggests that the 
magnitude of individual eigenvalues of strain would simply
scale as $\Sigma^{1/2}$. 
Fig.~\ref{fig:lam12_diss}a shows the conditional
average of first two eigenvalues,
and confirms this expectation
(the third eigenvalue, which has the largest
magnitude, can be obtained
via the incompressibility condition).   
It can also be seen that $\lambda_2$ is always positive,
but does not scale as $\Sigma^{1/2}$ for weak strain events
($\Sigma \tau_K^2 < 1$), and instead has a larger exponent.
This can be explained by realizing that when the magnitude of 
strain approaches zero, 
$\lambda_2$ would also approach zero, 
due to strong cancellation between $\lambda_1$ and
$\lambda_3$. This expectation is verified in 
Fig.~\ref{fig:lam12_diss}b,
which shows the quantity $\beta$ \cite{Ashurst87, BBP2020}, defined as:
\begin{align}
\beta = 
\sqrt{6} \frac{\lambda_2}{\sqrt{\lambda_1^2 + \lambda_2^2 + \lambda_3^2}}
= \sqrt{12} \frac{ \lambda_2}{\Sigma^{1/2}}
\label{eq:def_beta}
\end{align}
conditioned on $\Sigma$. It can be observed
that $\beta\to0$ when $\Sigma\to0$,
and only for $\Sigma\tau_K^2 > 1$,
it becomes a constant 
(provided the $\re$ is sufficiently high).
It is worth noting that this overall trend 
for $\lambda_2$ also explains the behavior
of strain self-amplification term in
Fig.~\ref{fig:trs3-str}a for the region $\Sigma \tau_K^2 < 1$.

\begin{figure}
\begin{center}
\includegraphics[width=0.46\textwidth]{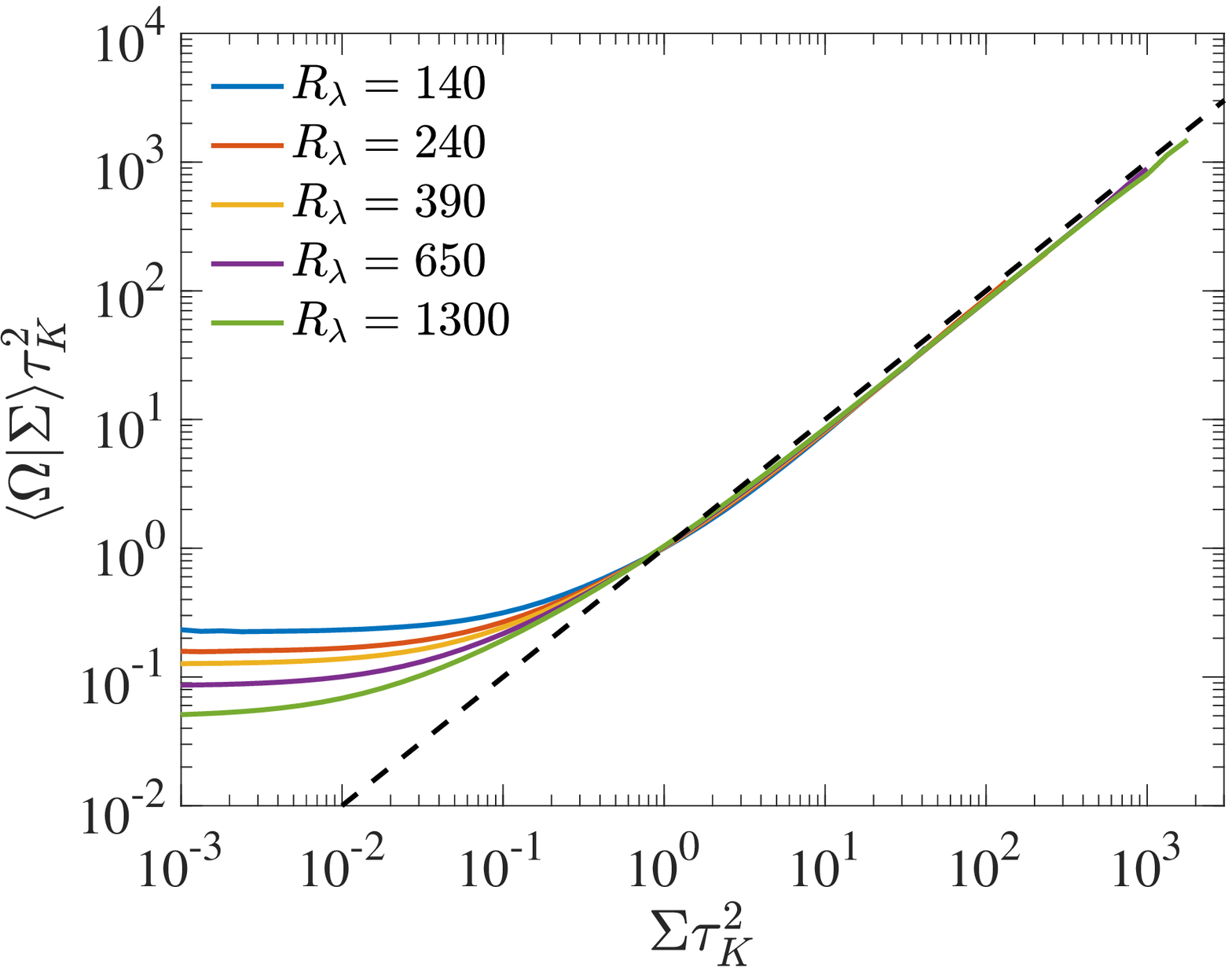} \ \ \ \ \
\includegraphics[width=0.46\textwidth]{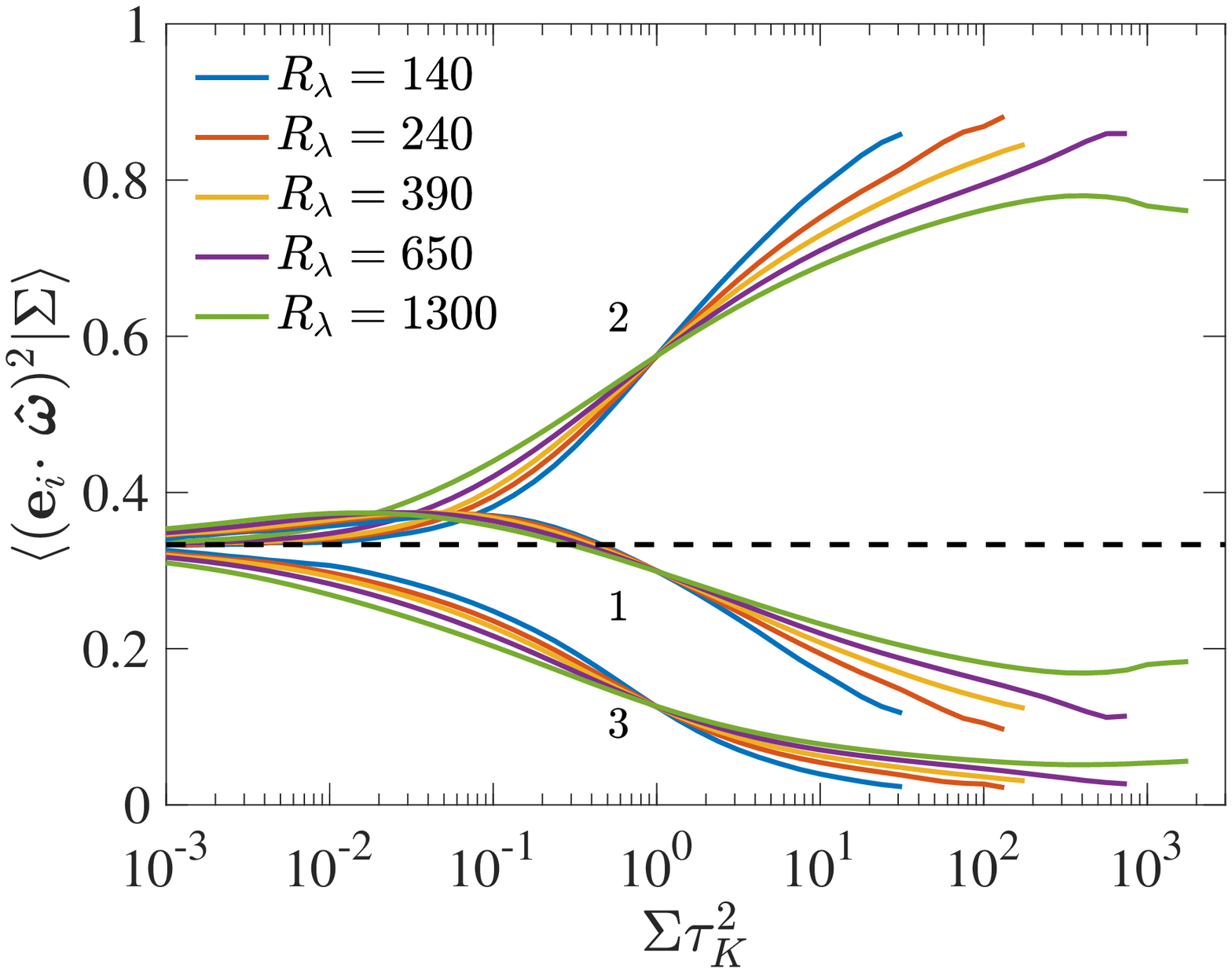} \\
\includegraphics[width=0.46\textwidth]{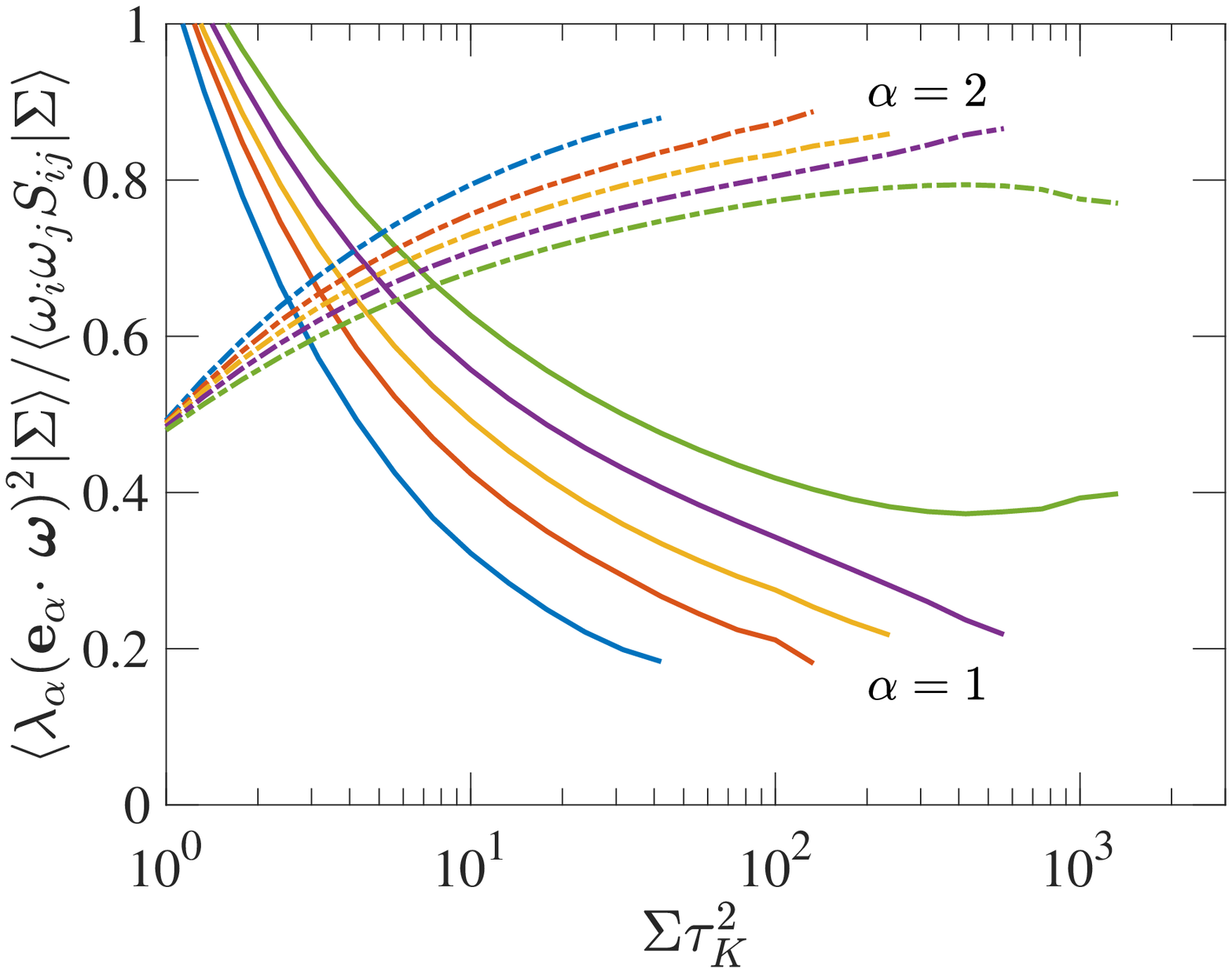}
\caption{
Conditional expectations (given $\Sigma$) 
of (a) enstrophy, $\Omega$, 
(b) second moment of alignment cosines between vorticity 
and eigenvectors of strain,
and (c) the relative contribution to the vortex stretching term
from the first two eigendirections of strain, for various $\re$.
The dashed line in (a) corresponds to a power-law of $\Sigma^1$. 
The dashed line in (b) is at $1/3$, corresponding to a
uniform distribution of alignment cosines 
(indicating lack of any preferential alignment).
}
\label{fig:vort}
\end{center}
\end{figure}

The vortex stretching term can be expressed in the eigenframe
of strain as:
\begin{align}
\omega_i \omega_j S_{ij} 
= \lambda_i (\mathbf{e}_i \cdot {\ww})^2 
= \Omega \lambda_i (\mathbf{e}_i \cdot \hat{\ww})^2 \ ,
\label{eq:wsw}
\end{align}
with $\hat{\ww} = \ww /|\ww|$, 
highlighting the importance of 
the alignment of vorticity with 
strain-eigenvectors (along with magnitude
of vorticity and strain) in determining
the efficacy of 
vortex stretching (or strain depletion in this case) 
\cite{Ashurst87, Tsi2009}.
%The vortex stretching term {therefore%}
%involves the magnitude of
%vorticity and also the various alignments {between vorticity
%and the strain eigenvectors,}
%as indicated in Eq.~\eqref{eq:wsw}.
The conditional expectation of enstrophy 
is shown in Fig.~\ref{fig:vort}a,
exhibiting the same qualitative behavior
as $\omega_i\omega_jS_{ij}$
in Fig.~\ref{fig:trs3-str}a. 
For intense strain events ($\Sigma \tau_K^2 > 1$),
we find that $\langle \Omega |\Sigma \rangle \sim \Sigma $,
suggesting a simple causal relation that intense
strain produces equally intense vorticity
(as anticipated from vortex stretching). 
%{The proportionality coefficient between $\langle \Omega | \Sigma \rangle $ 
%and $\Sigma$, $c_\Omega$, is slightly less than $1$:
%$c_\Omega \approx 0.8$.} \\
On the other hand, for weak strain events
($\Sigma \tau_K^2 < 1$), the conditional
average is constant, suggesting a
lack of correlation between
strain and vorticity \cite{Tsi2009, BBP2020}
(which is also reflected in the behavior
of $\omega_i\omega_jS_{ij}$
in Fig.~\ref{fig:trs3-str}a). 

%{Need to revisit from here till the
%end of the subsection.}
%The correspondence between results in
%Fig.~\ref{fig:vort}a and 
%Fig.~\ref{fig:trs3-str}a naively suggests that role
%of alignment cosines (between vorticity and strain)
%is likely not very significant for overall
%vortex stretching. However, this is not
%true, as the role of alignments would still be
%reflected in the overall magnitude
%of the term.

Figure~\ref{fig:vort}b shows the conditional
expectation of the second moment
of alignment cosines, i.e.,   
$\langle (\ei \cdot \hat{\ww})^2 | \Sigma\rangle$,
which are individually bounded between
$0$ and $1$, respectively for orthogonality 
and perfect alignment, and
are equal to $1/3$ for no preferential alignment
(corresponding to a uniform distribution of the cosine).
Additionally, the three alignment cosines
(for $i=1,2,3$) also 
add up to unity. 
Overall, the alignments follow the  
same trend as when conditioned on vorticity
(see Fig.3d in \cite{BBP2020}), i.e.,
in regions of intense strain ($\Sigma \tau_K^2 > 1$),
vorticity is strongly aligned with $\mathbf{e}_2$ 
and preferentially orthogonal to both $\mathbf{e}_{1,3}$
(more so with $\mathbf{e}_3$).
Whereas for $\Sigma \tau_K^2 \ll 1$, the alignments
approach $1/3$, reaffirming a lack of correlation
between strain and vorticity.
However, unlike when conditioned on vorticity (in \cite{BBP2020}),
the alignments in Fig.~\ref{fig:vort}b show a 
significant $\re$-dependence. The emergence of a plateau-like
behavior for $\re = 1300$ suggests 
an asymptotic state would likely be reached if $\re$ is
further increased. 
%(as evidenced by emergence of a plateau-like behavior
%for $\re=1300$).

%When $\Sigma \tau_K^2 \gg 1$, the values of 
%$\langle ( \ei \cdot \ew)^2 | \Sigma \rangle$ plausibly reach a plateau value,
%the vector $\ew$ being preferentially eigenvalue with $\mathbf{e}_2$, and 
%preferentially anti-aligned with $\mathbf{e}_3$, {and to some lesser extent with
%$\mathbf{e}_1$, as reflected by the values of}
%$\langle (\ew \cdot \ei)^2 | \Sigma \rangle$ being respectively 
%$ > 1/3$ {for $i=2$} and $ < 1/3$ {for $i = 3$ and $i = 1$}.
%At values of $\Sigma \tau_K^2 \lesssim 1$, on the contrary, no
%particular preferential alignment is observed, and 
%$\langle ( \ew \cdot \ei)^2 | \Sigma \rangle \approx 1/3$, 
%{for all indices $i$.}

Fig.~\ref{fig:vort}c shows the 
relative contributions of each eigenvalue
to the overall vortex stretching term.
We only  show contributions corresponding to first
and second eigenvalues, which are both positive).
The (negative) contribution for the third eigenvalue
is quite small, and can be evaluated by realizing that 
all three contributions add up to unity.
Interestingly, we notice that 
the contribution from the second eigenvalue is significantly 
stronger than that from the first eigenvalue, and accounts
for most of vortex stretching. The difference between the two
gradually decreases with $\re$, but
nevertheless, even at the highest $\re$ ($=1300$), the second
eigenvalue contributes to nearly $80\%$ of the net vortex stretching.
The results on alignments in Fig.~\ref{fig:vort}b, combined with these
indicate a strong structural difference between regions of intense 
strain and vorticity.
In regions of intense vorticity, even though vorticity is
strongly aligned with the second eigenvector,
the first eigenvalue contributes more significantly
to overall vortex stretching \cite{BBP2020}.
This difference can be explained by realizing
that the relative magnitude of 
$\lambda_2$ itself is significantly
smaller in regions of intense vorticity 
(compared to regions of intense strain) \cite{BBP2020}.

From a structural point of view, the above results
are consistent with the notion
that intense vorticity is arranged in tube-like
structures, whereas intense strain is arranged in
sheet-like structures \cite{moisy04}.
In both scenarios, vorticity has the
propensity to align with second eigenvector
of strain. However, for the case of vortex tubes, the
corresponding magnitude of second eigenvalue
is significantly smaller \cite{BBP2020}.
We will discuss more about this later in \S~3(c).

\subsection{Role of pressure Hessian}

\begin{figure}[h]
\begin{center}
\includegraphics[width=0.46\textwidth]{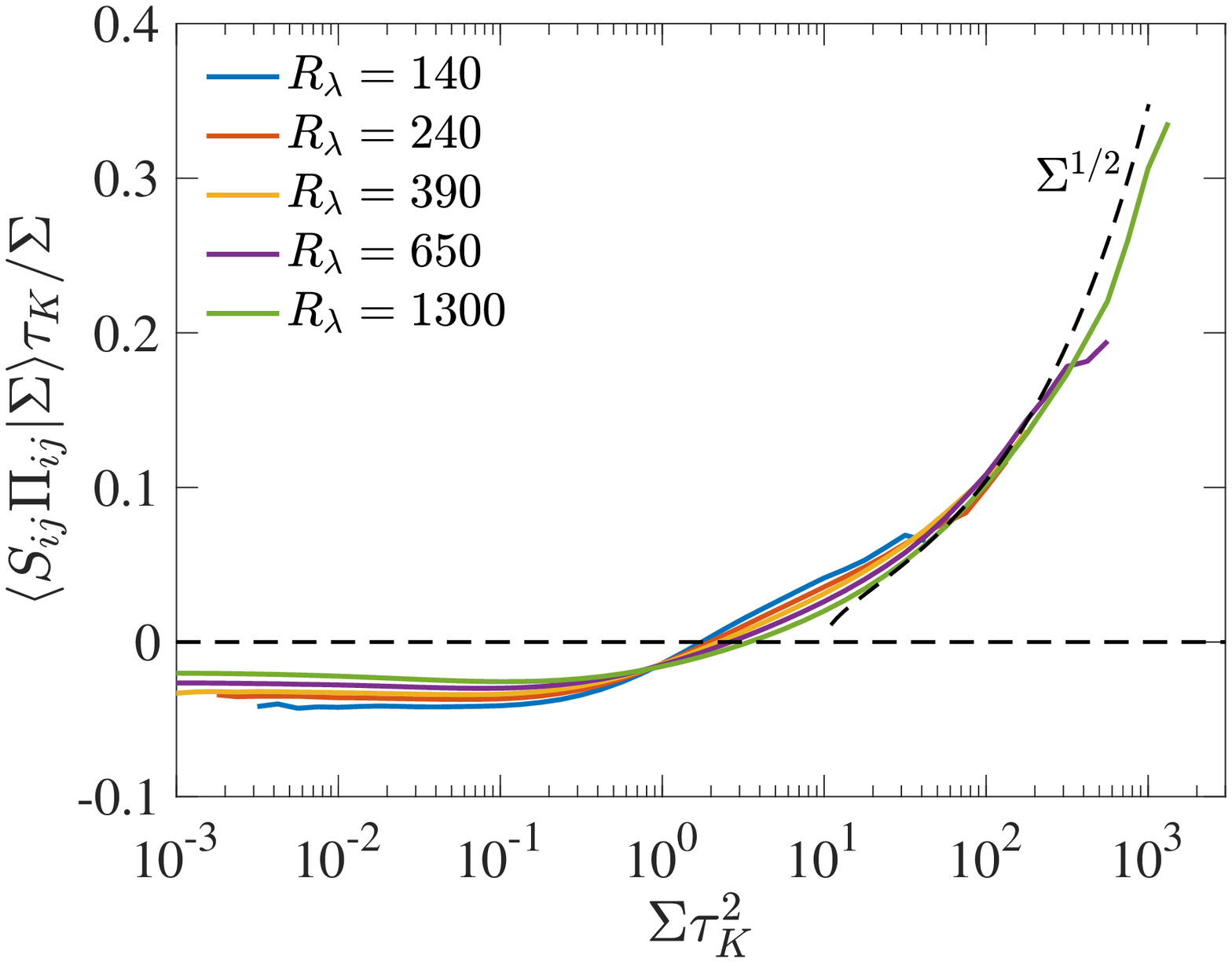} \ \ \ \ \ 
\includegraphics[width=0.46\textwidth]{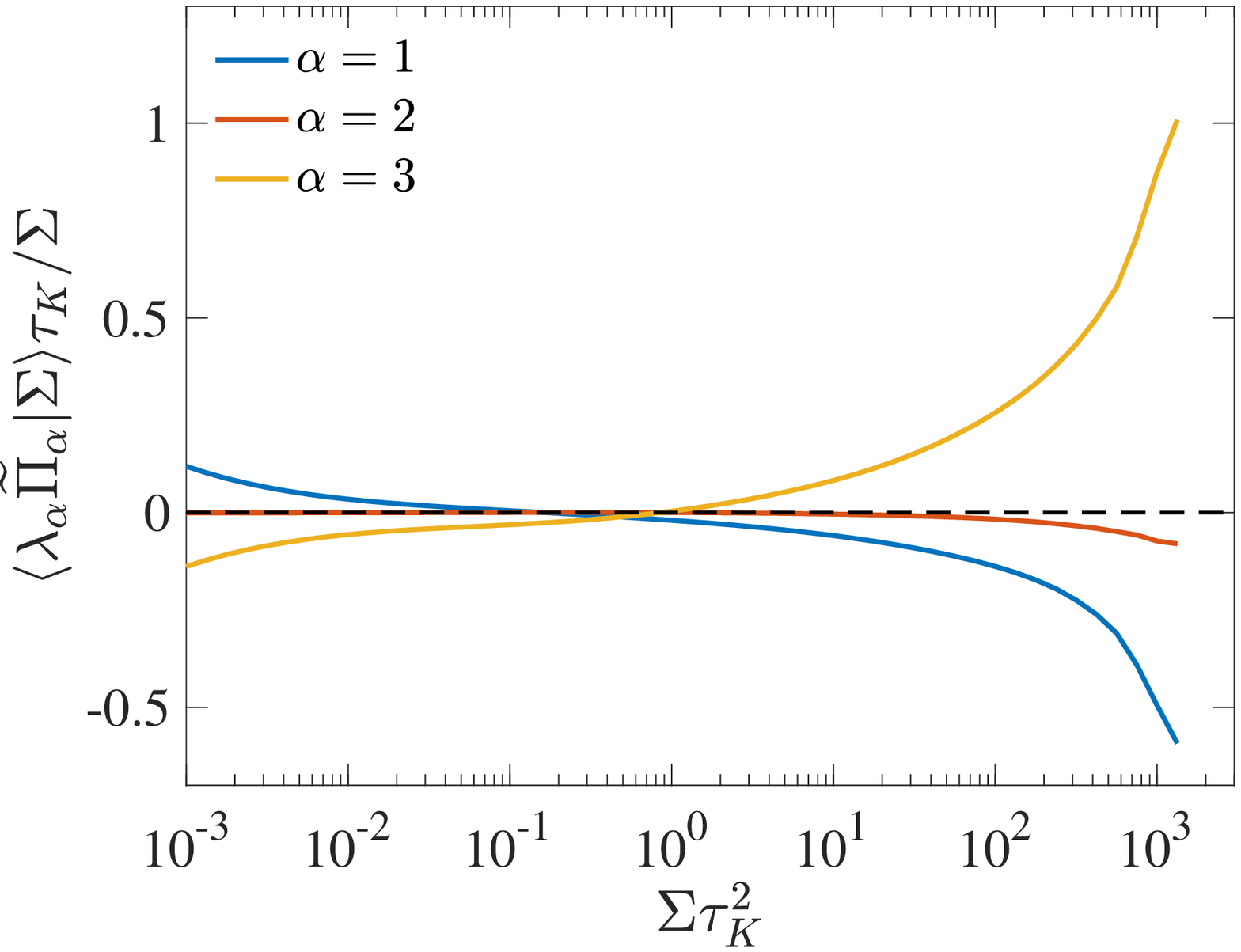}
\caption{
Conditional expectations (given $\Sigma$) 
of (a) strain and pressure-Hessian
correlation for various $\re$, and  
(b) the individual contributions
in the eigenframe of strain, as defined
by Eq.~\eqref{eq:sp3}, for only $\re=1300$. 
The expectations are once again divided by $\Sigma$,
and all quantities are non-dimensionalized by $\tau_K$.
}
\label{fig:Hp_s}
\end{center}
\end{figure}

We next consider the contribution of pressure Hessian to generation 
of strain. To this end, Fig.~\ref{fig:Hp_s}a shows 
the conditional average $\langle S_{ij} \Pi_{ij } | \Sigma \rangle$,
once again divided by $\Sigma$ %{and made dimensionless by multiplying by $\tau_K$.}
for convenience.
Since the corresponding unconditional average
is zero, the conditional average cannot keep the 
same sign for all values of $\Sigma$.
Figure~\ref{fig:Hp_s}a shows that
for events stronger than the mean 
($\Sigma \tau_K^2 \gtrsim 1$) this quantity
is positive, and thus leads to depletion
of strain (due to the negative sign associated
with the term in Eq.~\eqref{eq:ds2dt}), 
and vice-versa for 
events weaker than the mean. Thus, the 
non-local pressure field on average acts to 
redistribute the strain fluctuations
towards its mean amplitude \cite{tsi99}. 
It should be further noted, 
that for intense strain, the
conditional average scales once again
as $\Sigma^{3/2}$, whereas
for weak strain events it scales as $\Sigma^1$
-- albeit with a much
smaller pre-factor (in both regimes) when compared 
to the strain self-amplification or vortex stretching terms.

While the overall contribution of the pressure field
is to drive strain fluctuations towards the mean
field, it is once again instructive to analyze the individual
contributions in the eigenframe of strain tensor
-- where the strain-pressure Hessian correlation can be 
rewritten as:
\begin{align}
S_{ij} \Pi_{ij} = \lambda_i  \widetilde{\Pi}_i
\label{eq:sp3}
\end{align}
where 
$\widetilde{\Pi}_\alpha = \mathbf{e}_\alpha^T \ \boldsymbol{\Pi} \ \mathbf{e}_\alpha$
is the projection of pressure Hessian tensor along
the eigenvector $\mathbf{e}_\alpha$ of the strain.
and repeated $\alpha$ does not imply summation
(a convention which we will adhere to henceforth).
Note that the eigenvalues of strain can also be defined
in a similar way:
$\lambda_\alpha = \mathbf{e}_\alpha^T \mathbf{S} \mathbf{e}_\alpha$.
We also introduce the eigenframe of the
pressure Hessian tensor, defined by eigenvalues
$\lambda_i^p$ (for $i=1,2,3$ and also
arranged in descending order) 
and corresponding eigenvectors $\mathbf{e}_i^p$, leading to:
\begin{align}
\label{eq:lipi}
\widetilde{\Pi}_i &= \lambda_j^p (\mathbf{e}_i \cdot \mathbf{e}_j^p)^2  \ , 
\ \ \ \ \ \text{and} \\ 
\label{eq:lipi2}
S_{ij} \Pi_{ij}   &= \lambda_i  \lambda_j^p (\mathbf{e}_i \cdot \mathbf{e}_j^p)^2  %\\
\end{align}
Note 
$\sum_\alpha \lambda_\alpha^p = \nabla^2 P/\rho = (\Omega -\Sigma)/2$
(utilizing Eq.~\eqref{eq:poisson}).
Thus, from the result in Fig.~\ref{fig:vort}, 
i.e., $\langle \Omega |\Sigma \rangle \simeq c \Sigma^1$
(with $c \lesssim 1$),
it follows that in regions of intense strain
the sum of three eigenvalues is overall small
in magnitude with a negative sign -- which in turn suggests 
a dominant role of $\lambda_3^p$, which we analyze next.

Figure~\ref{fig:Hp_s}b shows the breakup of individual 
contributions to strain-pressure Hessian correlation as given in
Eq.~\eqref{eq:lipi}, i.e., 
$\langle \lambda_\alpha \widetilde{\Pi}_\alpha |\Sigma \rangle$
(we recall again that no summation is implied over $\alpha$).
It is observed that the dominant positive contribution
comes from the third eigenvalue of strain
(and hence leads to depletion of strain),
whereas the other two contributions are negative 
(leading to amplification of strain). 
These trends can be simply explained 
from Eq.~\eqref{eq:lipi2} by assuming that the alignments
$(\ei \cdot \mathbf{e}^p_j)^2$ are all
$1/3$ (corresponding to lack of any 
no preferential alignment) -- leading to
$\lambda_\alpha \widetilde{\Pi_\alpha} = 
\lambda_\alpha \sum_\alpha \lambda_\alpha^p = \lambda_\alpha \nabla^2 P/\rho$.
Since $\nabla^2 P$ is slightly negative
for $\Sigma \tau_K^2 > 1$, it follows that
$\lambda_\alpha \widetilde{\Pi_\alpha}$ has the 
opposite sign as that of $\lambda_\alpha$,
consistent with Fig.~\ref{fig:Hp_s}b.

\begin{figure}[h]
\begin{center}
\includegraphics[width=0.98\textwidth]{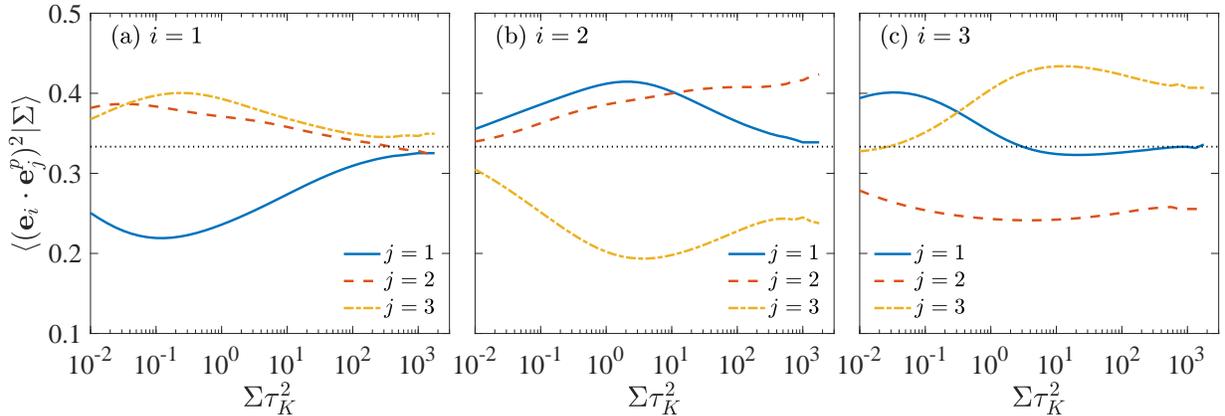}
\caption{
Conditional expectations (given $\Sigma$)
of the alignment cosines between eigenvectors
of strain and pressure Hessian, respectively
$\mathbf{e}_i$ and $\mathbf{e}_j^p$.
}
\label{fig:align}
\end{center}
\end{figure}

An important assumption in the argument above was the lack of
any preferential alignment
between the eigenvectors of strain and those of pressure Hessian.
Fig.~\ref{fig:align} shows the conditional second moments
of various alignment cosines,
i.e., $\langle (\mathbf{e}_i \cdot \mathbf{e}_j^p)^2 | \Sigma \rangle$. 
Note, similar to alignment cosines between vorticity and strain,
the alignment cosines for strain and pressure Hessian
for any fixed value of $i$ (or $j$) add up to unity.
Additionally, they are all individually bounded between $0$ and $1$,
for orthogonal and perfect alignment respectively; whereas
for non-preferential alignment the averages would be $1/3$
(corresponding to uniform random distribution).  
Fig.~\ref{fig:align} reveals that the deviation 
of all the alignments from $1/3$ is very small,
therefore excluding any strong alignments between 
the two sets of eigenvectors.
This is to be contrasted with the strong alignment observed
between vorticity and strain in Fig.~\ref{fig:vort}b.

%Fig.~\ref{fig:align} shows
%the quantities $\langle (\mathbf{e}_i \cdot \mathbf{e}^P_j)^2 | \Sigma \rangle$
%for $i = 1 $ (panel a), $i = 2$ (panel b) and $i = 3 $ (panel c), \AlP{and
%demonstrates that }
%$\mathbf{e}_2$ mildly aligns with $\mathbf{e}^P_2$,
%and correspondingly, mildly anti-aligns with $\mathbf{e}^P_3$, with no 
%particular alignment with $\mathbf{e}^P_1$. 
%Similarly, 
%$\mathbf{e}_3$ mildly aligns with $\mathbf{e}^P_3$,
%and mildly anti-aligns with $\mathbf{e}^P_2$, with again, no clear alignment
%with $\mathbf{e}^P_1$.  
%Finally, \AlP{the} overall mean values of the alignments, 
%$\langle (\mathbf{e}_i \cdot \mathbf{e}^P_j )^2 \rangle$, are indicated
%by an \AlP{upward ($ j = 1$), rightward ($j = 2$) and downward ($j = 3$)}
%pointing triangles in Fig.~\ref{fig:align}. These values
%are very close to the conditional values for $\Sigma \tau_K^2 = 0.6$: 
%$\langle ( \mathbf{e}_i \cdot \mathbf{e}^P_j )^2 \rangle \approx 
%\langle ( \mathbf{e}_i \cdot \mathbf{e}^P_j )^2 | \Sigma \approx 0.6/\tau_K^2) \rangle$.

\subsection{Budget of nonlinear terms and strain decomposition}
\label{sec:strain_decomp}

\begin{figure}[h]
\begin{center}
\includegraphics[width=0.46\textwidth]{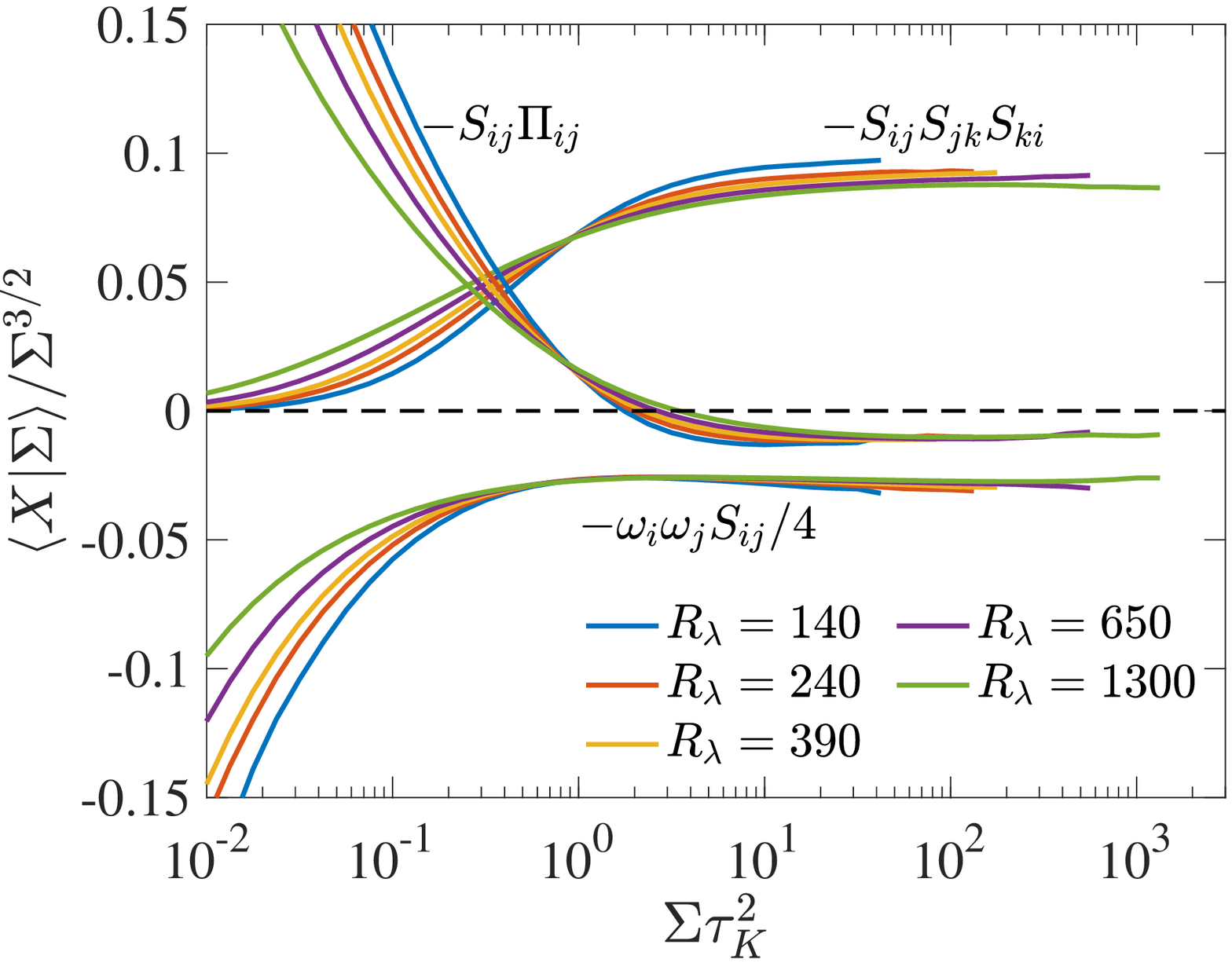} \ \ \ \ \ 
\includegraphics[width=0.46\textwidth]{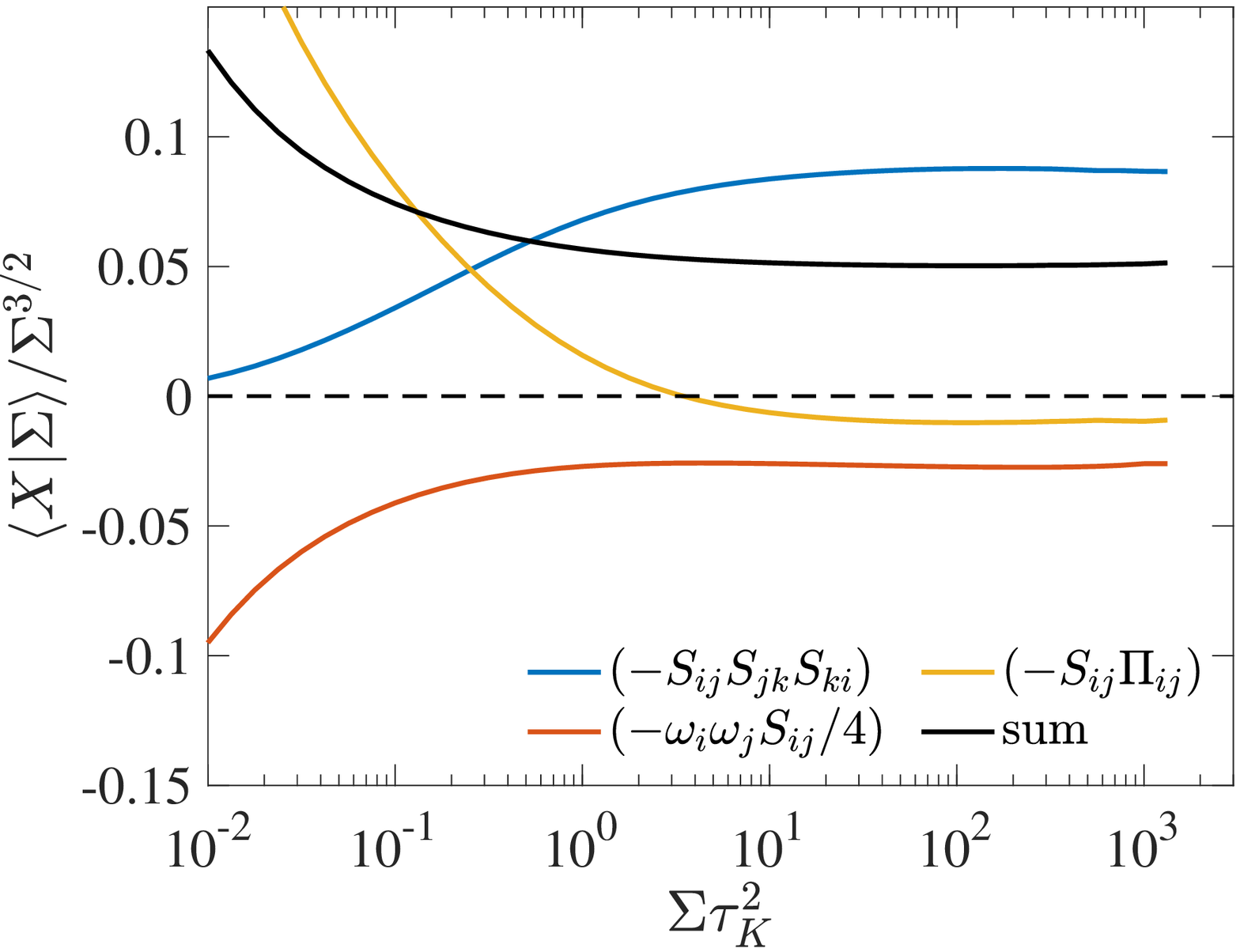}
\caption{
(a) Conditional expectations (given $\Sigma$) of
various nonlinear (inviscid) terms on the r.h.s.
of Eq.~\eqref{eq:ds2dt}, for various $\re$.
All quantities are normalized by $\Sigma^{3/2}$
revealing a plateau like behavior for $\Sigma \tau_K^2 >1$.
The same quantities are shown in (b) for $\re = 1300$, 
to highlight the combined contributions of the terms.
}
\label{fig:str_prod}
\end{center}
\end{figure}

Following upon the results in previous subsections,
Fig.~\ref{fig:str_prod}a compares the contributions 
of various nonlinear (inviscid) terms on the r.h.s. of 
Eq.~\eqref{eq:ds2dt} (note that the
viscous term is simply the negative of the net contribution 
of all the inviscid terms).
All the terms are now normalized 
by $\Sigma^{3/2}$, and clearly show
a plateau for $\Sigma \tau_K^2 > 1$. As expected, 
the dominant positive
contribution comes from the strain self-amplification term,
whereas the vortex stretching term is negative 
and significantly smaller in magnitude.
The contribution from pressure Hessian term is also negative
for $\Sigma \tau_K^2 \gtrsim 1$ and even smaller
in magnitude. 
To get more insight on the balance of terms in Eq.~\eqref{eq:ds2dt}, 
Fig.~\ref{fig:str_prod}b  shows the data only for $\re = 1300$,
including the resulting sum between various terms.
 
While the overall contribution of various inviscid
terms in generating intense strain, as illustrated in 
Fig.~\ref{fig:str_prod}, was relatively
straightforward, a more complex picture emerges
when considering amplification of individual
eigenvalues of strain. To this end, 
we consider the transport
equation for each eigenvalue
\cite{nomura:1998,Carbone:20b}:
\begin{align}
\frac{D \lambda_\alpha}{Dt} = 
- \lambda_\alpha^2 
+ \frac{\Omega}{4} \left[ 1 - (\mathbf{e}_\alpha \cdot \hat{\ww})^2 \right] 
-  \widetilde{\Pi}_{\alpha}  
+  {\rm viscous ~term}
\label{eq:dlambda_i}
\end{align} 
Multiplying both sides by $\lambda_\alpha$ leads
to the equation:
\begin{align}
\frac{1}{2} \frac{D \lambda_\alpha^2}{Dt} = 
- \lambda_\alpha^3 
+ \frac{1}{4} \Omega \lambda_\alpha \left[ 1 - (\mathbf{e}_\alpha \cdot \hat{\ww})^2 \right]
-  \lambda_\alpha \widetilde{\Pi}_{\alpha} 
+ {\rm viscous ~ terms}.
\label{eq:dlambda_sq_i}
\end{align} 
providing the individual breakups
for Eq.~\eqref{eq:ds2dt}, since 
$\Sigma = 2\sum_\alpha \lambda_\alpha^2$. 
Note that the individual eigenvalues now 
have a direct contribution from 
vorticity, from the term
$\Omega \lambda_\alpha$, which sums up to zero in 
Eq.~\eqref{eq:ds2dt} due to incompressibility.
This leads to a more involved interplay between strain 
self-amplification 
and vortex stretching at the level of 
individual eigenvalues, than for the 
total strain in Fig.~\ref{fig:str_prod}.

\begin{figure}
\begin{center}
\hspace{-0.5cm}
\includegraphics[width=0.98\textwidth]{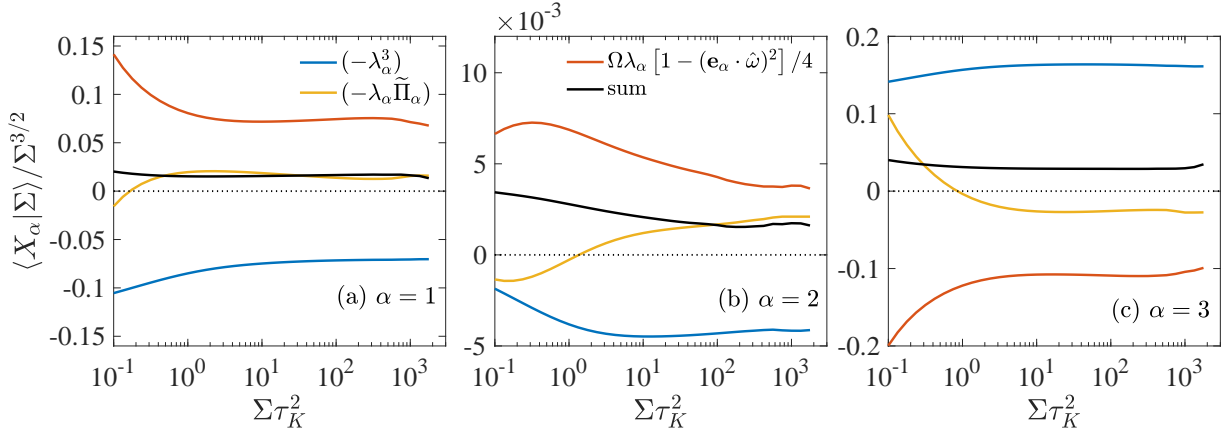} 
\caption{
Conditional expectations of the various inviscid terms 
in Eq.~\eqref{eq:dlambda_sq_i} for 
(a) $i = 1$, (b) $i = 2$, and (c) $i= 3$
(note that repeated indices do not imply summation). 
All quantities have been normalized by $\Sigma^{3/2}$. 
The legend in (a) and (b) is common to all figures.
}
\label{fig:Dlambda_i_sq}
\end{center}
\end{figure}

Since the first two eigenvalues of strain, $\lambda_1$
and $\lambda_2$, are positive, it follows that
the self-amplification term  $-\lambda_\alpha^3$,
leads to depletion instead of actual amplification;
whereas the contribution due to vortex stretching
is overall positive and leads to amplification 
(since $1 - (\mathbf{e}_\alpha \cdot \hat{\ww})^2 > 0$).
In contrast, for $\lambda_3$,
which is negative, the amplification originates from $- \lambda_3^3$, 
and vortex stretching leads to depletion. 
The sign of the pressure Hessian term, 
$-\lambda_\alpha \widetilde{\Pi}_\alpha$
will also be the same as that of $\lambda_\alpha$
(as shown in Fig.~\ref{fig:Hp_s}),
and thus would amplify $\lambda_1$ and $\lambda_2$,
but deplete $\lambda_3$.
These expectations are all qualitatively 
confirmed in Fig.~\ref{fig:Dlambda_i_sq}, which shows the
conditional averages of various
terms for each eigenvalue, conditioned
on $\Sigma$ and normalized by $\Sigma^{3/2}$.
Note, the individual contributions shown in 
Fig.~\ref{fig:Dlambda_i_sq}a-c sum up to the 
terms shown in Fig.~\ref{fig:str_prod}b.
Quantitatively, we find two main trends. 
For the case of $\lambda_1$ and $\lambda_2$,
in Fig.~\ref{fig:Dlambda_i_sq}a and b respectively,
the contributions from self-amplification and
vortex stretching terms approximately
cancel each other (for $\Sigma \tau_K^2 > 1$), 
and the net nonlinear amplification almost entirely
results from the pressure Hessian term.
For $\lambda_3$, in Fig.~\ref{fig:Dlambda_i_sq}c,
there is significant cancellation between
the self-amplification and vortex stretching terms
(and the pressure Hessian term now aids in depletion),
but the self-amplification overall dominates.

Thus, the following picture of strain amplification emerges.
The self-amplification of strain only occurs
along the third eigendirection, i.e. via self-compressive
motion, whereas the same mechanism depletes the first
two eigenvalues. On the other hand, as strain acts to amplify
vorticity (via vortex stretching), the feedback 
leads to amplification of the first
two eigenvalues (which could further aid in
vorticity amplification), but leads to depletion
of the third eigenvalue. The net result is such that
these two effects nearly balance each other 
for the first two eigenvalues, but the self-amplification
prevails for the third eigenvalue. Thus, overall these terms
only act to make $\lambda_3$ more negative (and thus produce
strong compressive motion). 
In this context, the pressure Hessian term acts to 
deplete the third
eigenvalue, and in turn amplify the first two, 
qualitatively producing a similar effect as 
vortex stretching.

These results clearly demonstrate that the 
generation of intense strain differs
in crucial aspects compared to 
generation of intense vorticity
\cite{Tsi2009, BBP2020}. 
While vortex stretching solely enables
generation of intense vorticity,
it also acts to deplete strain at the same 
spatial location.
This suggests that the local maximas of vorticity
and strain are never likely to colocated,
which has been confirmed
in DNS \cite{moisy04, elsinga17, BPBY2019} and also
corroborated by a simple vortex tube 
calculation \cite{moffatt94}.
At the same time, regions of intense
strain and vorticity also have
to be sufficiently `nearby' and
correlated, since strain and vorticity
are coupled via the Biot-Savart relation
\cite{ham_pof08,BPB2020}.  
This is also consistent with observations
from DNS, which show that intense strain
is arranged in sheet-like structures 
neighboring tube-like regions of 
intense vorticity \cite{BPBY2019}.  
Finally, these structural differences are in turn
consistent with how vorticity-strain
correlations differ between regions of intense
strain and vorticity. Thus, analyzing the non-local
relation between strain and vorticity could
be vital to understand their amplification \cite{BPB2020, BP2021} 
and also could provide a quantitative reasoning
as to why vorticity is more intermittent than strain \cite{BPBY2019}.

\section{Conclusion}
\label{sec:concl}

In this work, we have utilized
a massive DNS database of stationary isotropic
turbulence with Taylor-scale Reynolds number
in the range $140-1300$ to analyze
the nonlinear mechanisms
responsible for generation of extreme events of 
energy dissipation (and hence strain rate),
identified by $\Sigma = 2S_{ij} S{ij}$, where
$S_{ij}$ is the strain-rate tensor.
We have investigated the
three nonlinear processes involved in the 
transport equation for $\Sigma$
(see Eq.~\eqref{eq:ds2dt}), viz., the
strain self-amplification, vortex stretching
and strain-pressure Hessian correlation, by 
analyzing their statistics
conditioned on $\Sigma$.
We find that
the overall amplification of strain comes from
the strain-self amplification term only, 
% self-amplification term 
%produces the necessary amplification (as expected),
whereas the other two terms act to deplete 
intense strain events.
Remarkably, the dependence of various conditional averages 
on $\Sigma$ follows a 
simple dimensional consideration.
%all the conditional
%averages scale as $\Sigma^{3/2}$,
%which follows from a simple dimensional consideration.

%{Decomposing $\Sigma$ as a sum over the three eigenvalues of strain,
%$\lambda_i$, ranked in decreasing order, and studying the nonlinear
%amplification of each individual terms reveals}
The three mechanisms are further analyzed in the 
eigenbasis of strain tensor, defined by its 
eigenvalues $\lambda_i$ and eigenvectors $\mathbf{e}_i$ 
(for $i=1,2,3$), revealing 
a more complex picture.
Since $\lambda_1$ is always positive and $\lambda_2$ 
is positive on average, it follows that the self-amplification
term in fact leads to depletion for these eigenvalues,
whereas the vortex stretching and pressure Hessian terms
lead to amplification. Surprisingly, the self-amplification
and vortex stretching terms  cancel each other and the net
amplification is solely provided by the pressure Hessian term.
In contrast, the behavior of all these terms for
the third eigenvalue (which is always negative) 
is similar to that of total strain,
revealing that extreme events of strain result
from strong self-compressive action.
Our results are consistent with 
the notion that regions of intense strain are
arranged in sheet-like structures,
in the vicinity of, but never colocated with
regions of tube-like intense vorticity 
\cite{moffatt94, moisy04, elsinga17, BPBY2019}.
In this context, analyzing the non-local relation
between strain and vorticity would be crucial
in understanding their amplification 
\cite{ham_pof08, BPB2020, BP2021}
and could also shed light on the long standing
mystery of why vorticity is more intermittent
than strain.

\begin{table}[h]
\centering
\resizebox{\linewidth}{!}{%
    \begin{tabular}{l|l||l|l||l|l||l|l||l|l}
\hline
    $X$       & $c_X \Sigma^{1/2}$ & $X$        & $c_X \Sigma^1$ & $X$       & $c_X \Sigma^{3/2}$ & $X$             & $c_X \Sigma^{3/2}$ & $X$       & $c_X \Sigma^{3/2}$ \\ 
\hline
\hline
    $\lambda_1$ & $0.412$    & $\lambda_1^2$ & $0.173$    & $\lambda_1^3$            & $0.074$    & $\lambda_1 (\mathbf{e}_1 \cdot \ww)^2$ & $0.042$    & $\lambda_1 \widetilde{\Pi}_1$ & $0.013$        \\
    $\lambda_2$ & $0.138$    & $\lambda_2^2$ & $0.024$    & $\lambda_2^3$            & $0.004$    & $\lambda_2 (\mathbf{e}_2 \cdot \ww)^2$ & $0.090$    & $\lambda_2 \widetilde{\Pi}_2$ & $0.002$        \\
    $\lambda_3$ & $-0.550$   & $\lambda_3^2$ & $0.303$    & $\lambda_3^3$            & $-0.167$   & $\lambda_3 (\mathbf{e}_3 \cdot \ww)^2$ & $-0.019$   & $\lambda_3 \widetilde{\Pi}_3$ & $-0.024$        \\
                &            & $\Omega$      & $0.856$    & $S_{ij} S_{jk} S_{ki}$   & $-0.089$   & $\omega_i \omega_j S_{ij}$             & $0.113$    & $S_{ij} \Pi_{ij}$             & $-0.010$   \\
\hline
    \end{tabular}
}
%\end{table}
%\begin{table}[h]
%\centering
%    \begin{tabular}{c||cccc||ccccc}
%\hline
% $X$ &   $\Omega $  & $\lambda_1^2 $  & $ \lambda_2^2 $  & $ \lambda_3^2 $ & 
%$\omega_i \omega_j S_{ij} $ &
%$S_{ij} S_{jk} S_{ki} $ &
%$ \lambda_1^3  $ &
%$ \lambda_2^3  $ &
%$ \lambda_3^3 $  \\
%\hline
%$c_X$ &    $ 0.8$ & $ 0.164$ & $0.024$ & $0.29$ & $0.105$ &  $-0.083$ & $0.065$ & $0.004$ & $ -0.154$  \\
%\hline
%    \end{tabular}
\caption{Pre-factors $c_X$ for the asymptotic scaling of 
various conditional statistics satisfying 
$\langle X | \Sigma \rangle \approx c_X \Sigma^{p}$
(in the range $\Sigma / \langle \Sigma \rangle > 1$),
such that $X/\Sigma^{p}$ is dimensionless
$\lambda_i$ and $\mathbf{e}_i$ are the eigenvalues
and eigenvectors of strain tensor respectively.
$\ww$ is the vorticity and $\Omega=\omega_i \omega_i$.
$\Pi_{ij}$ is the pressure Hessian tensor, and
$\widetilde{\Pi}_i$ is its projection along $\mathbf{e}_i$.
The results are obtained from the $\re=1300$ data.
}
\label{tab:prefac}
\end{table}

Finally, we note that the conditional statistics 
investigated in this work have very simple power-law dependences on $\Sigma$
(in the region $\Sigma / \langle \Sigma \rangle > 1$), 
as deducible from an elementary dimensional consideration.
We have listed all such relevant quantities in Table~\ref{tab:prefac},
which could be valuable in statistical modeling of 
energy dissipation rate, especially in PDF methods \cite{pope1994} 
-- an exercise left for future work.

\section*{Acknowledgements}
We are pleased to present this manuscript
in the collection honoring the momentous career
of Professor Uriel Frisch.
We gratefully acknowledge the Gauss Centre for Supercomputing e.V.
for providing computing time on the supercomputers JUQUEEN and
JUWELS at J\"ulich Supercomputing Centre (JSC),
where the simulations reported in this
paper were performed.

%\bibliography{large_grad}

\begin{thebibliography}{30}%
\makeatletter
\providecommand \@ifxundefined [1]{%
 \@ifx{#1\undefined}
}%
\providecommand \@ifnum [1]{%
 \ifnum #1\expandafter \@firstoftwo
 \else \expandafter \@secondoftwo
 \fi
}%
\providecommand \@ifx [1]{%
 \ifx #1\expandafter \@firstoftwo
 \else \expandafter \@secondoftwo
 \fi
}%
\providecommand \natexlab [1]{#1}%
\providecommand \enquote  [1]{``#1''}%
\providecommand \bibnamefont  [1]{#1}%
\providecommand \bibfnamefont [1]{#1}%
\providecommand \citenamefont [1]{#1}%
\providecommand \href@noop [0]{\@secondoftwo}%
\providecommand \href [0]{\begingroup \@sanitize@url \@href}%
\providecommand \@href[1]{\@@startlink{#1}\@@href}%
\providecommand \@@href[1]{\endgroup#1\@@endlink}%
\providecommand \@sanitize@url [0]{\catcode `\\12\catcode `\$12\catcode
  `\&12\catcode `\#12\catcode `\^12\catcode `\_12\catcode `\%12\relax}%
\providecommand \@@startlink[1]{}%
\providecommand \@@endlink[0]{}%
\providecommand \url  [0]{\begingroup\@sanitize@url \@url }%
\providecommand \@url [1]{\endgroup\@href {#1}{\urlprefix }}%
\providecommand \urlprefix  [0]{URL }%
\providecommand \Eprint [0]{\href }%
\providecommand \doibase [0]{http://dx.doi.org/}%
\providecommand \selectlanguage [0]{\@gobble}%
\providecommand \bibinfo  [0]{\@secondoftwo}%
\providecommand \bibfield  [0]{\@secondoftwo}%
\providecommand \translation [1]{[#1]}%
\providecommand \BibitemOpen [0]{}%
\providecommand \bibitemStop [0]{}%
\providecommand \bibitemNoStop [0]{.\EOS\space}%
\providecommand \EOS [0]{\spacefactor3000\relax}%
\providecommand \BibitemShut  [1]{\csname bibitem#1\endcsname}%
\let\auto@bib@innerbib\@empty
%</preamble>
\bibitem [{\citenamefont {Frisch}(1995)}]{Frisch95}%
  \BibitemOpen
  \bibfield  {author} {\bibinfo {author} {\bibfnamefont {U.}~\bibnamefont
  {Frisch}},\ }\href@noop {} {\emph {\bibinfo {title} {Turbulence: the legacy
  of {Kolmogorov}}}}\ (\bibinfo  {publisher} {Cambridge University Press},\
  \bibinfo {address} {Cambridge},\ \bibinfo {year} {1995})\BibitemShut
  {NoStop}%
\bibitem [{\citenamefont {Sreenivasan}(1984)}]{sreeni84}%
  \BibitemOpen
  \bibfield  {author} {\bibinfo {author} {\bibfnamefont {K.~R.}\ \bibnamefont
  {Sreenivasan}},\ }\bibfield  {title} {\enquote {\bibinfo {title} {On the
  scaling of the turbulence energy dissipation rate},}\ }\href@noop {}
  {\bibfield  {journal} {\bibinfo  {journal} {Phys.~Fluids}\ }\textbf {\bibinfo
  {volume} {27}},\ \bibinfo {pages} {1048--1051} (\bibinfo {year}
  {1984})}\BibitemShut {NoStop}%
\bibitem [{\citenamefont {Kaneda}\ \emph {et~al.}(2003)\citenamefont {Kaneda},
  \citenamefont {Ishihara}, \citenamefont {Yokokawa}, \citenamefont {Itakura},\
  and\ \citenamefont {Uno}}]{kaneda03}%
  \BibitemOpen
  \bibfield  {author} {\bibinfo {author} {\bibfnamefont {Y.}~\bibnamefont
  {Kaneda}}, \bibinfo {author} {\bibfnamefont {T.}~\bibnamefont {Ishihara}},
  \bibinfo {author} {\bibfnamefont {M.}~\bibnamefont {Yokokawa}}, \bibinfo
  {author} {\bibfnamefont {K.}~\bibnamefont {Itakura}}, \ and\ \bibinfo
  {author} {\bibfnamefont {A.}~\bibnamefont {Uno}},\ }\bibfield  {title}
  {\enquote {\bibinfo {title} {Energy dissipation rate and energy spectrum in
  high resolution direct numerical simulations of turbulence in a periodic
  box},}\ }\href@noop {} {\bibfield  {journal} {\bibinfo  {journal}
  {Phys.~Fluids}\ }\textbf {\bibinfo {volume} {15}},\ \bibinfo {pages}
  {L21--L24} (\bibinfo {year} {2003})}\BibitemShut {NoStop}%
\bibitem [{\citenamefont {Meneveau}\ and\ \citenamefont
  {Sreenivasan}()}]{MS91}%
  \BibitemOpen
  \bibfield  {author} {\bibinfo {author} {\bibfnamefont {C.}~\bibnamefont
  {Meneveau}}\ and\ \bibinfo {author} {\bibfnamefont {K.~R.}\ \bibnamefont
  {Sreenivasan}},\ }\bibfield  {title} {\enquote {\bibinfo {title} {The
  multifractal nature of turbulent energy dissipation},}\ }\href@noop {}
  {\bibfield  {journal} {\bibinfo  {journal} {J.~Fluid Mech.}\ }\textbf
  {\bibinfo {volume} {224}},\ \bibinfo {pages} {429--484}}\BibitemShut
  {NoStop}%
\bibitem [{\citenamefont {Buaria}\ \emph {et~al.}(2019)\citenamefont {Buaria},
  \citenamefont {Pumir}, \citenamefont {Bodenschatz},\ and\ \citenamefont
  {Yeung}}]{BPBY2019}%
  \BibitemOpen
  \bibfield  {author} {\bibinfo {author} {\bibfnamefont {D.}~\bibnamefont
  {Buaria}}, \bibinfo {author} {\bibfnamefont {A.}~\bibnamefont {Pumir}},
  \bibinfo {author} {\bibfnamefont {E.}~\bibnamefont {Bodenschatz}}, \ and\
  \bibinfo {author} {\bibfnamefont {P.~K.}\ \bibnamefont {Yeung}},\ }\bibfield
  {title} {\enquote {\bibinfo {title} {Extreme velocity gradients in turbulent
  flows},}\ }\href@noop {} {\bibfield  {journal} {\bibinfo  {journal} {New
  J.~Phys.}\ }\textbf {\bibinfo {volume} {21}},\ \bibinfo {pages} {043004}
  (\bibinfo {year} {2019})}\BibitemShut {NoStop}%
\bibitem [{\citenamefont {Sreenivasan}\ and\ \citenamefont
  {Antonia}(1997)}]{Sreeni97}%
  \BibitemOpen
  \bibfield  {author} {\bibinfo {author} {\bibfnamefont {K.~S.}\ \bibnamefont
  {Sreenivasan}}\ and\ \bibinfo {author} {\bibfnamefont {R.~A.}\ \bibnamefont
  {Antonia}},\ }\bibfield  {title} {\enquote {\bibinfo {title} {The
  phenomenology of small-scale turbulence},}\ }\href@noop {} {\bibfield
  {journal} {\bibinfo  {journal} {Annu.~Rev.~Fluid~Mech.}\ }\textbf {\bibinfo
  {volume} {29}},\ \bibinfo {pages} {435--77} (\bibinfo {year}
  {1997})}\BibitemShut {NoStop}%
\bibitem [{\citenamefont {Tsinober}(2009)}]{Tsi2009}%
  \BibitemOpen
  \bibfield  {author} {\bibinfo {author} {\bibfnamefont {A.}~\bibnamefont
  {Tsinober}},\ }\href@noop {} {\emph {\bibinfo {title} {An Informal Conceptual
  Introduction to Turbulence}}}\ (\bibinfo  {publisher} {Springer},\ \bibinfo
  {address} {Berlin},\ \bibinfo {year} {2009})\BibitemShut {NoStop}%
\bibitem [{\citenamefont {Buaria}\ \emph {et~al.}(2015)\citenamefont {Buaria},
  \citenamefont {Sawford},\ and\ \citenamefont {Yeung}}]{BSY.2015}%
  \BibitemOpen
  \bibfield  {author} {\bibinfo {author} {\bibfnamefont {D.}~\bibnamefont
  {Buaria}}, \bibinfo {author} {\bibfnamefont {B.~L.}\ \bibnamefont {Sawford}},
  \ and\ \bibinfo {author} {\bibfnamefont {P.~K.}\ \bibnamefont {Yeung}},\
  }\bibfield  {title} {\enquote {\bibinfo {title} {Characteristics of backward
  and forward two-particle relative dispersion in turbulence at different
  {R}eynolds numbers},}\ }\href@noop {} {\bibfield  {journal} {\bibinfo
  {journal} {Phys. Fluids}\ }\textbf {\bibinfo {volume} {27}},\ \bibinfo
  {pages} {105101} (\bibinfo {year} {2015})}\BibitemShut {NoStop}%
\bibitem [{\citenamefont {Buaria}\ \emph {et~al.}(2021)\citenamefont {Buaria},
  \citenamefont {Clay}, \citenamefont {Sreenivasan},\ and\ \citenamefont
  {Yeung}}]{BCSY2021a}%
  \BibitemOpen
  \bibfield  {author} {\bibinfo {author} {\bibfnamefont {D.}~\bibnamefont
  {Buaria}}, \bibinfo {author} {\bibfnamefont {M.~P.}\ \bibnamefont {Clay}},
  \bibinfo {author} {\bibfnamefont {K.~R.}\ \bibnamefont {Sreenivasan}}, \ and\
  \bibinfo {author} {\bibfnamefont {P.~K.}\ \bibnamefont {Yeung}},\ }\bibfield
  {title} {\enquote {\bibinfo {title} {Small-scale isotropy and ramp-cliff
  structures in scalar turbulence},}\ }\href@noop {} {\bibfield  {journal}
  {\bibinfo  {journal} {Phys.~Rev.~Lett.}\ }\textbf {\bibinfo {volume} {126}},\
  \bibinfo {pages} {034504} (\bibinfo {year} {2021})}\BibitemShut {NoStop}%
\bibitem [{\citenamefont {Pitsch}\ and\ \citenamefont
  {Steiner}(2000)}]{Pitsch2000}%
  \BibitemOpen
  \bibfield  {author} {\bibinfo {author} {\bibfnamefont {H.}~\bibnamefont
  {Pitsch}}\ and\ \bibinfo {author} {\bibfnamefont {H.}~\bibnamefont
  {Steiner}},\ }\bibfield  {title} {\enquote {\bibinfo {title} {Scalar mixing
  and dissipation rate in large-eddy simulations of non-premixed turbulent
  combustion},}\ }\href@noop {} {\bibfield  {journal} {\bibinfo  {journal}
  {Proc. Combust. Inst.}\ }\textbf {\bibinfo {volume} {28}},\ \bibinfo {pages}
  {41--49} (\bibinfo {year} {2000})}\BibitemShut {NoStop}%
\bibitem [{\citenamefont {Hamlington}\ \emph {et~al.}(2011)\citenamefont
  {Hamlington}, \citenamefont {Poludnenko},\ and\ \citenamefont
  {Oran}}]{ham_pof11}%
  \BibitemOpen
  \bibfield  {author} {\bibinfo {author} {\bibfnamefont {P.~E.}\ \bibnamefont
  {Hamlington}}, \bibinfo {author} {\bibfnamefont {A.~Y.}\ \bibnamefont
  {Poludnenko}}, \ and\ \bibinfo {author} {\bibfnamefont {E.~S.}\ \bibnamefont
  {Oran}},\ }\bibfield  {title} {\enquote {\bibinfo {title} {Interactions
  between turbulence and flames in premixed reacting flows},}\ }\href@noop {}
  {\bibfield  {journal} {\bibinfo  {journal} {Physics of Fluids}\ }\textbf
  {\bibinfo {volume} {23}},\ \bibinfo {pages} {125111} (\bibinfo {year}
  {2011})}\BibitemShut {NoStop}%
\bibitem [{\citenamefont {Tennekes}\ and\ \citenamefont {Lumley}(1972)}]{tl72}%
  \BibitemOpen
  \bibfield  {author} {\bibinfo {author} {\bibfnamefont {H.}~\bibnamefont
  {Tennekes}}\ and\ \bibinfo {author} {\bibfnamefont {J.~L.}\ \bibnamefont
  {Lumley}},\ }\href@noop {} {\emph {\bibinfo {title} {A First Course in
  Turbulence}}}\ (\bibinfo  {publisher} {The MIT Press},\ \bibinfo {year}
  {1972})\BibitemShut {NoStop}%
\bibitem [{\citenamefont {Sundaram}\ and\ \citenamefont
  {Collins}(1997)}]{collins97}%
  \BibitemOpen
  \bibfield  {author} {\bibinfo {author} {\bibfnamefont {S.}~\bibnamefont
  {Sundaram}}\ and\ \bibinfo {author} {\bibfnamefont {L.~R.}\ \bibnamefont
  {Collins}},\ }\bibfield  {title} {\enquote {\bibinfo {title} {Collision
  statistics in an isotropic particle-laden turbulent suspension. {Part} 1.
  {Direct} numerical simulations},}\ }\href@noop {} {\bibfield  {journal}
  {\bibinfo  {journal} {J.~Fluid Mech.}\ }\textbf {\bibinfo {volume} {335}},\
  \bibinfo {pages} {75--109} (\bibinfo {year} {1997})}\BibitemShut {NoStop}%
\bibitem [{\citenamefont {Hamlington}\ \emph {et~al.}(2008)\citenamefont
  {Hamlington}, \citenamefont {Schumacher},\ and\ \citenamefont
  {Dahm}}]{ham_pof08}%
  \BibitemOpen
  \bibfield  {author} {\bibinfo {author} {\bibfnamefont {P.~E.}\ \bibnamefont
  {Hamlington}}, \bibinfo {author} {\bibfnamefont {J.}~\bibnamefont
  {Schumacher}}, \ and\ \bibinfo {author} {\bibfnamefont {W.~J.~A.}\
  \bibnamefont {Dahm}},\ }\bibfield  {title} {\enquote {\bibinfo {title}
  {Direct assessment of vorticity alignment with local and nonlocal strain
  rates in turbulent flows},}\ }\href@noop {} {\bibfield  {journal} {\bibinfo
  {journal} {Phys.~Fluids}\ }\textbf {\bibinfo {volume} {20}},\ \bibinfo
  {pages} {111703} (\bibinfo {year} {2008})}\BibitemShut {NoStop}%
\bibitem [{\citenamefont {Buaria}\ \emph
  {et~al.}(2020{\natexlab{a}})\citenamefont {Buaria}, \citenamefont {Pumir},\
  and\ \citenamefont {Bodenschatz}}]{BPB2020}%
  \BibitemOpen
  \bibfield  {author} {\bibinfo {author} {\bibfnamefont {D.}~\bibnamefont
  {Buaria}}, \bibinfo {author} {\bibfnamefont {A.}~\bibnamefont {Pumir}}, \
  and\ \bibinfo {author} {\bibfnamefont {E.}~\bibnamefont {Bodenschatz}},\
  }\bibfield  {title} {\enquote {\bibinfo {title} {Self-attenuation of extreme
  events in {Navier-Stokes} turbulence},}\ }\href@noop {} {\bibfield  {journal}
  {\bibinfo  {journal} {Nat. Commun.}\ }\textbf {\bibinfo {volume} {11}},\
  \bibinfo {pages} {5852} (\bibinfo {year} {2020}{\natexlab{a}})}\BibitemShut
  {NoStop}%
\bibitem [{\citenamefont {Buaria}\ and\ \citenamefont {Pumir}(2021)}]{BP2021}%
  \BibitemOpen
  \bibfield  {author} {\bibinfo {author} {\bibfnamefont {D.}~\bibnamefont
  {Buaria}}\ and\ \bibinfo {author} {\bibfnamefont {A.}~\bibnamefont {Pumir}},\
  }\bibfield  {title} {\enquote {\bibinfo {title} {Nonlocal amplification of
  intense vorticity in turbulent flows},}\ }\href@noop {} {\bibfield  {journal}
  {\bibinfo  {journal} {arXiv:2106.14370}\ } (\bibinfo {year}
  {2021})}\BibitemShut {NoStop}%
\bibitem [{\citenamefont {Buaria}\ \emph
  {et~al.}(2020{\natexlab{b}})\citenamefont {Buaria}, \citenamefont
  {Bodenschatz},\ and\ \citenamefont {Pumir}}]{BBP2020}%
  \BibitemOpen
  \bibfield  {author} {\bibinfo {author} {\bibfnamefont {D.}~\bibnamefont
  {Buaria}}, \bibinfo {author} {\bibfnamefont {E.}~\bibnamefont {Bodenschatz}},
  \ and\ \bibinfo {author} {\bibfnamefont {A.}~\bibnamefont {Pumir}},\
  }\bibfield  {title} {\enquote {\bibinfo {title} {Vortex stretching and
  enstrophy production in high {Reynolds} number turbulence},}\ }\href@noop {}
  {\bibfield  {journal} {\bibinfo  {journal} {Phys.~Rev.~Fluids}\ }\textbf
  {\bibinfo {volume} {5}},\ \bibinfo {pages} {104602} (\bibinfo {year}
  {2020}{\natexlab{b}})}\BibitemShut {NoStop}%
\bibitem [{\citenamefont {Carbone}\ and\ \citenamefont
  {Bragg}(2020)}]{carbone20}%
  \BibitemOpen
  \bibfield  {author} {\bibinfo {author} {\bibfnamefont {M.}~\bibnamefont
  {Carbone}}\ and\ \bibinfo {author} {\bibfnamefont {A.~D.}\ \bibnamefont
  {Bragg}},\ }\bibfield  {title} {\enquote {\bibinfo {title} {Is vortex
  stretching the main cause of the turbulent energy cascade?}}\ }\href@noop {}
  {\bibfield  {journal} {\bibinfo  {journal} {J.~Fluid Mech.}\ }\textbf
  {\bibinfo {volume} {883}},\ \bibinfo {pages} {R2} (\bibinfo {year}
  {2020})}\BibitemShut {NoStop}%
\bibitem [{\citenamefont {Johnson}(2021)}]{johnson21}%
  \BibitemOpen
  \bibfield  {author} {\bibinfo {author} {\bibfnamefont {P.~L.}\ \bibnamefont
  {Johnson}},\ }\bibfield  {title} {\enquote {\bibinfo {title} {On the role of
  vorticity stretching and strain self-amplification in the turbulence energy
  cascade},}\ }\href@noop {} {\bibfield  {journal} {\bibinfo  {journal}
  {arXiv:2102.06844}\ } (\bibinfo {year} {2021})}\BibitemShut {NoStop}%
\bibitem [{\citenamefont {Moisy}\ and\ \citenamefont
  {Jim{\'e}nez}(2004)}]{moisy04}%
  \BibitemOpen
  \bibfield  {author} {\bibinfo {author} {\bibfnamefont {F.}~\bibnamefont
  {Moisy}}\ and\ \bibinfo {author} {\bibfnamefont {J.}~\bibnamefont
  {Jim{\'e}nez}},\ }\bibfield  {title} {\enquote {\bibinfo {title} {Geometry
  and clustering of intense structures in isotropic turbulence},}\ }\href@noop
  {} {\bibfield  {journal} {\bibinfo  {journal} {J.~Fluid Mech.}\ }\textbf
  {\bibinfo {volume} {513}},\ \bibinfo {pages} {111--133} (\bibinfo {year}
  {2004})}\BibitemShut {NoStop}%
\bibitem [{\citenamefont {Elsinga}\ \emph {et~al.}(2017)\citenamefont
  {Elsinga}, \citenamefont {Ishihara}, \citenamefont {Goudar}, \citenamefont
  {Da~Silva},\ and\ \citenamefont {Hunt}}]{elsinga17}%
  \BibitemOpen
  \bibfield  {author} {\bibinfo {author} {\bibfnamefont {G.~E.}\ \bibnamefont
  {Elsinga}}, \bibinfo {author} {\bibfnamefont {T.}~\bibnamefont {Ishihara}},
  \bibinfo {author} {\bibfnamefont {M.~V.}\ \bibnamefont {Goudar}}, \bibinfo
  {author} {\bibfnamefont {C.~B.}\ \bibnamefont {Da~Silva}}, \ and\ \bibinfo
  {author} {\bibfnamefont {J.~C.~R.}\ \bibnamefont {Hunt}},\ }\bibfield
  {title} {\enquote {\bibinfo {title} {The scaling of straining motions in
  homogeneous isotropic turbulence},}\ }\href@noop {} {\bibfield  {journal}
  {\bibinfo  {journal} {J.~Fluid Mech.}\ }\textbf {\bibinfo {volume} {829}},\
  \bibinfo {pages} {31--64} (\bibinfo {year} {2017})}\BibitemShut {NoStop}%
\bibitem [{\citenamefont {Jim{\'e}nez}\ \emph {et~al.}(1993)\citenamefont
  {Jim{\'e}nez}, \citenamefont {Wray}, \citenamefont {Saffman},\ and\
  \citenamefont {Rogallo}}]{Jimenez93}%
  \BibitemOpen
  \bibfield  {author} {\bibinfo {author} {\bibfnamefont {J.}~\bibnamefont
  {Jim{\'e}nez}}, \bibinfo {author} {\bibfnamefont {A.~A.}\ \bibnamefont
  {Wray}}, \bibinfo {author} {\bibfnamefont {P.~G.}\ \bibnamefont {Saffman}}, \
  and\ \bibinfo {author} {\bibfnamefont {R.~S.}\ \bibnamefont {Rogallo}},\
  }\bibfield  {title} {\enquote {\bibinfo {title} {The structure of intense
  vorticity in isotropic turbulence},}\ }\href@noop {} {\bibfield  {journal}
  {\bibinfo  {journal} {J. Fluid Mech.}\ }\textbf {\bibinfo {volume} {255}}
  (\bibinfo {year} {1993})}\BibitemShut {NoStop}%
\bibitem [{\citenamefont {Moffatt}\ \emph {et~al.}(1994)\citenamefont
  {Moffatt}, \citenamefont {Kida},\ and\ \citenamefont {Ohkitani}}]{moffatt94}%
  \BibitemOpen
  \bibfield  {author} {\bibinfo {author} {\bibfnamefont {H.~K.}\ \bibnamefont
  {Moffatt}}, \bibinfo {author} {\bibfnamefont {S.}~\bibnamefont {Kida}}, \
  and\ \bibinfo {author} {\bibfnamefont {K.}~\bibnamefont {Ohkitani}},\
  }\bibfield  {title} {\enquote {\bibinfo {title} {Stretched vortices--the
  sinews of turbulence; {large-Reynolds-number} asymptotics},}\ }\href@noop {}
  {\bibfield  {journal} {\bibinfo  {journal} {J.~Fluid Mech.}\ }\textbf
  {\bibinfo {volume} {259}},\ \bibinfo {pages} {241--264} (\bibinfo {year}
  {1994})}\BibitemShut {NoStop}%
\bibitem [{\citenamefont {Buaria}\ and\ \citenamefont
  {Sreenivasan}(2020)}]{BS2020}%
  \BibitemOpen
  \bibfield  {author} {\bibinfo {author} {\bibfnamefont {D.}~\bibnamefont
  {Buaria}}\ and\ \bibinfo {author} {\bibfnamefont {K.~R.}\ \bibnamefont
  {Sreenivasan}},\ }\bibfield  {title} {\enquote {\bibinfo {title} {Dissipation
  range of the energy spectrum in high {Reynolds} number turbulence},}\
  }\href@noop {} {\bibfield  {journal} {\bibinfo  {journal}
  {Phys.~Rev.~Fluids}\ }\textbf {\bibinfo {volume} {5}},\ \bibinfo {pages}
  {092601(R)} (\bibinfo {year} {2020})}\BibitemShut {NoStop}%
\bibitem [{\citenamefont {Betchov}(1956)}]{Betchov56}%
  \BibitemOpen
  \bibfield  {author} {\bibinfo {author} {\bibfnamefont {R.}~\bibnamefont
  {Betchov}},\ }\bibfield  {title} {\enquote {\bibinfo {title} {An inequality
  concerning the production of vorticity in isotropic turbulence},}\
  }\href@noop {} {\bibfield  {journal} {\bibinfo  {journal} {J.~Fluid Mech.}\
  }\textbf {\bibinfo {volume} {1}},\ \bibinfo {pages} {497--504} (\bibinfo
  {year} {1956})}\BibitemShut {NoStop}%
\bibitem [{\citenamefont {Ashurst}\ \emph {et~al.}(1987)\citenamefont
  {Ashurst}, \citenamefont {Kerstein}, \citenamefont {Kerr},\ and\
  \citenamefont {Gibson}}]{Ashurst87}%
  \BibitemOpen
  \bibfield  {author} {\bibinfo {author} {\bibfnamefont {W.~T.}\ \bibnamefont
  {Ashurst}}, \bibinfo {author} {\bibfnamefont {A.~R.}\ \bibnamefont
  {Kerstein}}, \bibinfo {author} {\bibfnamefont {R.~M.}\ \bibnamefont {Kerr}},
  \ and\ \bibinfo {author} {\bibfnamefont {C.~H.}\ \bibnamefont {Gibson}},\
  }\bibfield  {title} {\enquote {\bibinfo {title} {Alignment of vorticity and
  scalar gradient with strain rate in simulated {Navier-Stokes} turbulence},}\
  }\href@noop {} {\bibfield  {journal} {\bibinfo  {journal} {Phys. Fluids}\
  }\textbf {\bibinfo {volume} {30}},\ \bibinfo {pages} {2343--2353} (\bibinfo
  {year} {1987})}\BibitemShut {NoStop}%
\bibitem [{\citenamefont {Tsinober}\ \emph {et~al.}(1999)\citenamefont
  {Tsinober}, \citenamefont {Ortenberg},\ and\ \citenamefont
  {Shtilman}}]{tsi99}%
  \BibitemOpen
  \bibfield  {author} {\bibinfo {author} {\bibfnamefont {A.}~\bibnamefont
  {Tsinober}}, \bibinfo {author} {\bibfnamefont {M.}~\bibnamefont {Ortenberg}},
  \ and\ \bibinfo {author} {\bibfnamefont {L.}~\bibnamefont {Shtilman}},\
  }\bibfield  {title} {\enquote {\bibinfo {title} {On depression of
  nonlinearity in turbulence},}\ }\href@noop {} {\bibfield  {journal} {\bibinfo
   {journal} {Phys. Fluids}\ }\textbf {\bibinfo {volume} {11}},\ \bibinfo
  {pages} {2291--2297} (\bibinfo {year} {1999})}\BibitemShut {NoStop}%
\bibitem [{\citenamefont {Nomura}\ and\ \citenamefont
  {Post}(1998)}]{nomura:1998}%
  \BibitemOpen
  \bibfield  {author} {\bibinfo {author} {\bibfnamefont {K.~K.}\ \bibnamefont
  {Nomura}}\ and\ \bibinfo {author} {\bibfnamefont {G.~K.}\ \bibnamefont
  {Post}},\ }\bibfield  {title} {\enquote {\bibinfo {title} {The structure and
  dynamics of vorticity and rate of strain in incompressible homogeneous
  turbulence},}\ }\href@noop {} {\bibfield  {journal} {\bibinfo  {journal} {J.
  Fluid Mech.}\ }\textbf {\bibinfo {volume} {377}},\ \bibinfo {pages} {65--97}
  (\bibinfo {year} {1998})}\BibitemShut {NoStop}%
\bibitem [{\citenamefont {Carbone}\ \emph {et~al.}(2020)\citenamefont
  {Carbone}, \citenamefont {Iovieno},\ and\ \citenamefont
  {Bragg}}]{Carbone:20b}%
  \BibitemOpen
  \bibfield  {author} {\bibinfo {author} {\bibfnamefont {M.}~\bibnamefont
  {Carbone}}, \bibinfo {author} {\bibfnamefont {M.}~\bibnamefont {Iovieno}}, \
  and\ \bibinfo {author} {\bibfnamefont {A.~D.}\ \bibnamefont {Bragg}},\
  }\bibfield  {title} {\enquote {\bibinfo {title} {Symmetry transformation and
  dimensionality reduction of the anistropic pressure hessian},}\ }\href@noop
  {} {\bibfield  {journal} {\bibinfo  {journal} {J.~Fluid Mech.}\ }\textbf
  {\bibinfo {volume} {900}},\ \bibinfo {pages} {A38} (\bibinfo {year}
  {2020})}\BibitemShut {NoStop}%
\bibitem [{\citenamefont {Pope}(1994)}]{pope1994}%
  \BibitemOpen
  \bibfield  {author} {\bibinfo {author} {\bibfnamefont {S.B.}\ \bibnamefont
  {Pope}},\ }\bibfield  {title} {\enquote {\bibinfo {title} {Lagrangian {PDF}
  methods for turbulent flows},}\ }\href@noop {} {\bibfield  {journal}
  {\bibinfo  {journal} {Annu.~Rev.~Fluid Mech.}\ }\textbf {\bibinfo {volume}
  {26}},\ \bibinfo {pages} {23--63} (\bibinfo {year} {1994})}\BibitemShut
  {NoStop}%
\end{thebibliography}


\end{document}